\documentclass[fleqn,usenatbib]{mnras}

\usepackage{newtxtext,newtxmath}

\usepackage[T1]{fontenc}

\DeclareRobustCommand{\VAN}[3]{#2}
\let\VANthebibliography\thebibliography
\def\thebibliography{\DeclareRobustCommand{\VAN}[3]{##3}\VANthebibliography}

\usepackage{graphicx}	
\usepackage{amsmath}	
\usepackage{subfigure}

\title[Revisiting the TeV flare of PKS~2155-304]{Revisiting the TeV flare of PKS~2155-304 in 2006
}

\author[Tan et al.]{
Hong-Bin Tan,$^{1,2}$
Ruo-Yu Liu,$^{1,2}$
Markus B{\"o}ttcher$^{3}$
\\
$^{1}$School of Astronomy and Space Science, Nanjing University, Nanjing 210023, China; {\color{blue} \rm ryliu@nju.edu.cn}\\
$^{2}$Key laboratory of Modern Astronomy and Astrophysics (Nanjing University), Ministry of Education, Nanjing 210023, People's Republic of China\\
$^{3}$Centre for Space Research, North-West University, Potchefstroom 2520, South Africa 
}

\date{Accepted 2024 February 21. Received 2024 January 31; in original form 2023 November 22}

\pubyear{2023}

\begin{document}
\label{firstpage}
\pagerange{\pageref{firstpage}--\pageref{lastpage}}
\maketitle

\begin{abstract}
Blazars, a subclass of active galactic nuclei (AGN), are known to be bright $\gamma$-ray sources, frequently exhibiting active (flaring) periods. The blazar PKS~2155-304 is a high synchrotron-peaked BL Lac object located at redshift $z=0.116$. On 2006 July 28, an extremely remarkable outburst of VHE $\gamma$-ray emission from this blazar was reported by the H.E.S.S. experiment, with an average flux more than 10 times the low-state level. The variability timescale of this extraordinary flare was as short as approximately 200~s. In order to guarantee the transparency of the emission region for TeV photons, the fast variability demands an extremely high Doppler factor $\delta_{\rm D}>50$ of the jet within the classical one-zone model, leading to the so-called ``Doppler factor crisis''. Here we demonstrate that the stochastic dissipation model, which is a multi-blob scenario for blazars, can self-consistently explain the giant TeV flares of PKS~2155-304 and the low-state emission before and after the flares, in terms of both multi-wavelength spectral and variability characteristics. The required Doppler factor in this model can be as low as 20, which is a reasonable and typical value for blazar jets. The obtained model parameters may shed some light on the physical properties of the relativistic jet.
\end{abstract}




\section{Introduction}\label{sec:intro}

Blazars are the most extreme subclass of active galactic nuclei (AGN) with a relativistic jet almost pointing to the observer \citep[][]{1995PASP..107..803U}. Blazars can be divided into flat spectrum radio quasars (FSRQs) and BL Lacertae objects (BL Lacs) based on whether there is strong broad line emission in the optical and ultraviolet band. The emission of a blazar is generally considered to be dominated by the non-thermal radiation of its relativistic jet. The broad band spectrum, ranging from the radio to the $\gamma$-ray band, displays a typical double-hump structure \citep[e.g.,][]{1998MNRAS.299..433F}. The low energy hump from radio to UV/X-rays is widely believed to be produced by the relativistic electrons through synchrotron radiation. On the other hand, the origin of the high energy component is generally believed to originate from the inverse Compton scattering of the synchrotron photons produced by the same electron population \citep[synchrotron-self Compton, SSC; e.g.,][]{1974ApJ...188..353J, 1996ApJ...461..657B} and/or external thermal soft photons from surrounding structures \citep[external Compton, EC; e.g.,][]{1993ApJ...416..458D, 1994ApJ...421..153S} such as broad line region (BLR) or dusty torus (DT), with possible additional contributions from hadronic processes \citep[e.g.,][]{2000NewA....5..377A, 2013ApJ...768...54B, 2022A&A...659A.184L, 2023MNRAS.518.2719P}. Another important feature of blazar emission is the intense temporal variability observed in various wavelengths, which provides insights into the microphysical processes in the jet. In particular, huge flux enhancements, known as flares, are observed from different blazars from time to time. A broad range of timescales of flares are reported, from month to years \citep[e.g.,][]{2011ApJS..192...12I, 2013A&A...553A.107C} down to hours \citep[e.g.][]{2011ApJ...738...25A, 2015ApJ...807...79H} or even minutes \citep[e.g.,][]{2005ApJ...622..160X, 2007ApJ...669..862A}. Due to light-travel-time arguments, such rapid outbursts must be caused by compact and powerful regions within the jet \citep[e.g.][]{1995ARA&A..33..163W, 2013A&A...557A..71R}. In contrast to these rapid flares, the long-term variability on month-to-year long timescales can be related to changes in the global parameters of the blazar jet \citep[e.g.,][]{2004A&A...419..913O}. 

PKS~2155-304 is a high synchrotron-peaked BL Lac with redshift $z=0.116$. It was discovered by the HEAO 1 X-ray satellite \citep{1979ApJ...234..810G, 1979ApJ...229L..53S} and it is one of the brightest extragalactic X-ray sources. It has been monitored across the electromagnetic spectrum for many years. $\gamma$-ray emission was detected by the EGRET detector aboard the Compton Gamma Ray Observatory satellite from 30 MeV to 10 GeV during 1994 November 15-29 \citep{1995ApJ...454L..93V}. Furthermore, it was detected by the University of Durham Mark 6 Telescope above 300 GeV in 1996 and 1997 \citep{1999ApJ...513..161C}, and was therefore identified as a TeV $\gamma$-ray emitter. This was confirmed by the observation of H.E.S.S. in 2002 and 2003, where significant $\gamma$-ray emission above 160 GeV was reported \citep{2005A&A...430..865A}. For a more comprehensive understanding of this TeV blazar, many multiwavelength observation campaigns were conducted and resulted in many simultaneous, high-quality broad band data sets \citep[e.g.,][]{2005A&A...442..895A}.

On 2006 July 28 (MJD~53944), an extreme outburst of VHE $\gamma$-ray emission of PKS 2155-304 was reported by H.E.S.S. \citep{2007ApJ...664L..71A}. During this extraordinary outburst, the VHE flux exhibited amplitudes ranging from 0.65 to 15.1 times the Crab Nebula flux (denoted by $I_{\rm Crab}$) above 200\,GeV, with an average flux $7I_{\rm Crab}$ --- more than 10 times the typically observed value of $0.15I_{\rm Crab}$. Well-resolved sub-flares exhibited variability timescales as short as 200~s, implying that the size of the emission region must be exceptionally small, $R \lesssim 5.4 \times 10^{13} \, \delta_1$~cm (where $\delta_1 = \delta_{\rm D}/10$ and $\delta_{\rm D} = 1/\sqrt{1 - \beta_{\Gamma} \, \cos\theta_{\rm obs}}$ is the Doppler factor, with $\Gamma$ being the bulk Lorentz factor of emitting material moving along the jet at an angle $\theta_{\rm obs}$ with respect to our line of sight). Unfortunately, there were no simultaneous multiwavelength observations available during that night. In the night of 2006 July 29 (MJD~53945), another significant $\gamma$-ray outburst took place 44 hours after the first outburst \citep{2009A&A...502..749A}. The $\gamma$-ray flux reached 11 times the Crab flux above 400\,GeV of the major flare and several main sub-flares were detected within a few hours. During the second outburst, Chandra, Swift \citep{2007ApJ...657L..81F}, RXTE and the Bronberg Observatory were simultaneously monitoring this source, among which the Chandra and Bronberg Observatory observations covered most of the flare period. Except for these two outbursts, the source experienced a relatively low flux state in the night between the two dramatic events \citep{2012A&A...539A.149H}. These observations revealed both the long-term evolution and the fast variations of this TeV blazar.

A widely used and classical model to study the spectrum and variability of blazars is the one-zone leptonic model \citep[e.g.,][]{1996ApJ...463..555I, 1997A&A...320...19M, 2008ApJ...686..181F}. It is assumed that all the non-thermal radiation of a blazar is generated from a compact spherical plasma blob filled with high-energy electrons inside the jet. For a flare with a variability timescale of $\Delta t$, the radius of the blob $R$ need be smaller than $\delta_{\rm D} c \Delta t / (1 + z)$, since otherwise the variability would be smeared out due to the light-travel-time difference of photons emitted from different parts of the blob. On the other hand, if the radiation zone (or the blob) is too compact, the photon density of the synchrotron radiation emitted by electrons would be very high and hence produce a high opacity for high-energy gamma-ray photons via the pair production process. Therefore, in the exceptional $\gamma$-ray outburst of PKS 2155-304 on 2006 July 28, if the whole emission is produced in a single compact region, an unusually high Doppler factor $\delta_{\rm D}>50$ is needed to ensure that the VHE $\gamma$-ray photons are able to escape the blob \citep{2008MNRAS.384L..19B}. This value is much higher than the one inferred from radio observations, and this is often referred to as the ``Doppler factor crisis'' \citep{2010ApJ...722..197L}. Some multi-zone models are established to reduce the photon density in the emission region to avoid the strong absorption effect, such as the two-zone time-dependent SSC model containing a long-term component dominating the X-ray emission and a flaring component accounting for the higher-energy (IC) emission \citep{2008MNRAS.390..371K}, and the jet/shock-in-a-jet model \citep[][see also \citealt{1985ApJ...298..114M} ]{2009MNRAS.395L..29G, 2011MNRAS.413..333N} which may account for the demanded high Doppler factor of the radiation zone.

Recently, the so-called stochastic dissipation model, a multi-blob framework, has been proposed to explain the multi-wavelength spectral energy distributions (SEDs), light curves (LCs) and polarization of blazars \citep{2022PhRvD.105b3005W, 2023MNRAS.526.5054L}. It assumes that numerous blobs are generated along the blazar jet following a certain probability distribution. The radiation of the blazar would be the sum of each blob in the jet. Within this framework, \citet{2023MNRAS.526.5054L} showed that in the low-state of a blazar, the predicted gamma-ray opacity is smaller than that in the one-zone model, and minute-scale variability in the LC in the TeV band can be reproduced with the flux varying by about a factor of 2. However, it is not clear yet whether or not a giant minute-scale TeV flare such as that observed in PKS~2155-304 can be explained in the model. In this work, following the stochastic dissipation model, we study the outburst events caused by several blobs with strong dissipation, and try to reproduce the multi-wavelength spectrum and the minute-to-hour scale variability in the case of PKS 2155-304. In particular, we will search for a solution with a relatively modest Doppler factor, aiming to avoid the Doppler factor crisis. The rest of this paper is structured as follows. In Section~\ref{sec:model}, the multi-blob framework is briefly introduced. In Section~\ref{sec:app}, we apply our model to the two outbursts of PKS~2155-304 observed on 28 July 2006 (MJD~53944) and 30 July 2006 (MJD~53946). We discuss the results in Section~\ref{sec:discussion}, and the conclusions are presented in Section~\ref{sec:conclusion}. Throughout the paper, we adopt the $\Lambda$CDM cosmology with $H_{\rm{0}}={69.6\, \rm{km~s^{-1} Mpc^{-1}}}$, $\Omega_{\rm{m}}=0.29$, $\Omega_{\rm{\Lambda}}=0.71$.

\begin{figure*}
\centering
\subfigure{
\includegraphics[width=0.9\columnwidth]{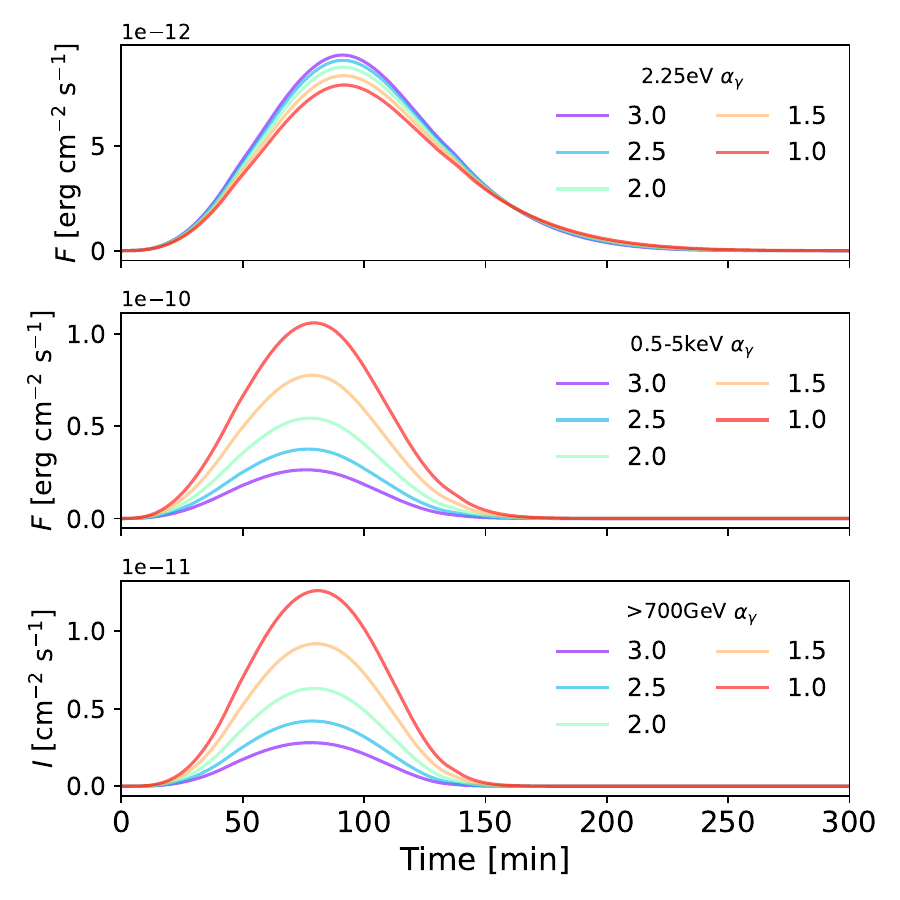}
}\hspace{-5mm}
\quad
\subfigure{
\includegraphics[width=0.9\columnwidth]{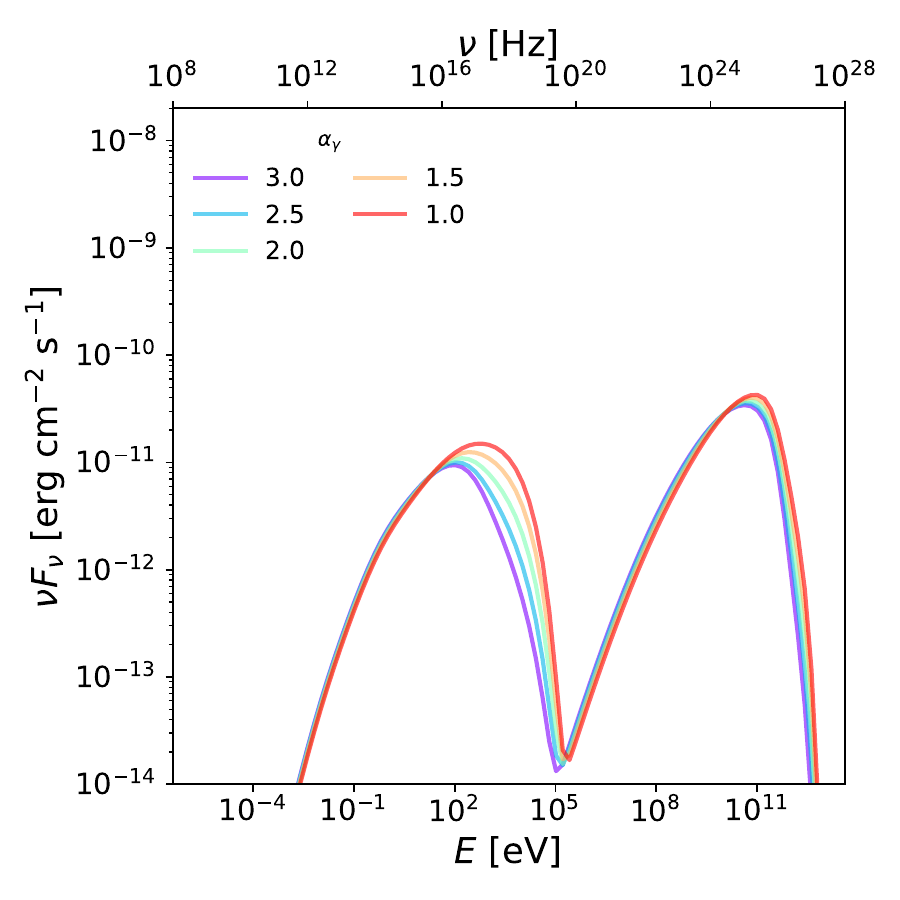}
}\hspace{-5mm}
\quad
\caption{The calculated LCs (left panel) and average SEDs (right panel) of a segment with different $\alpha_{\gamma}$ varying from 1.0 to 3.0. The redshift is assumed as $z=0.116$, and other parameters are set as $r_{\rm seg}=0.02$pc, $\rm \delta_D=20$, $N_{\rm blob}=480$ indicating a filling factor $\lambda\approx60\%$, $B=0.08$G, $L^{\prime}_{\rm inj}=1.0\times 10^{40} \rm erg\ s^{-1}$, $s=2.0$, $\gamma^{\prime}_{\rm min}=5.0\times 10^{3}$, $\gamma^{\prime}_{\rm max,1}=1.0\times 10^{5}$, $\gamma^{\prime}_{\rm max,2}=1.2\times 10^{6}$. For the sake of simplicity, all blobs in this segment are assumed to be generated at the same time. From the figure, a smaller $\alpha_{\gamma}$ can result in a larger $\gamma$-ray and X-ray flux accompanying a lower optical flux, and a harder spectrum on the right side of the peaks as interpreted in Eq.(\ref{eq:alpha_gamma}). Here the EBL absorption is taken into account for VHE $\gamma$-ray emission.}
\label{fig:test_alpha_gamma}
\end{figure*}

\begin{figure*}
\centering
\subfigure{
\includegraphics[width=0.9\columnwidth]{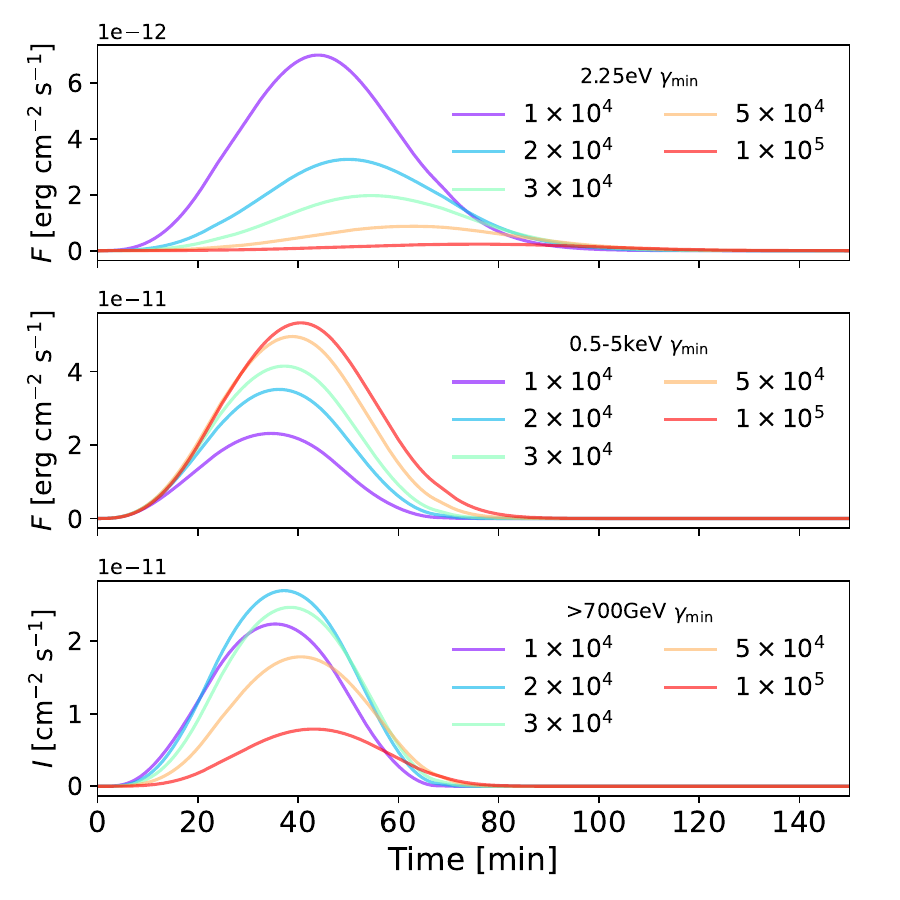}
}\hspace{-5mm}
\quad
\subfigure{
\includegraphics[width=0.9\columnwidth]{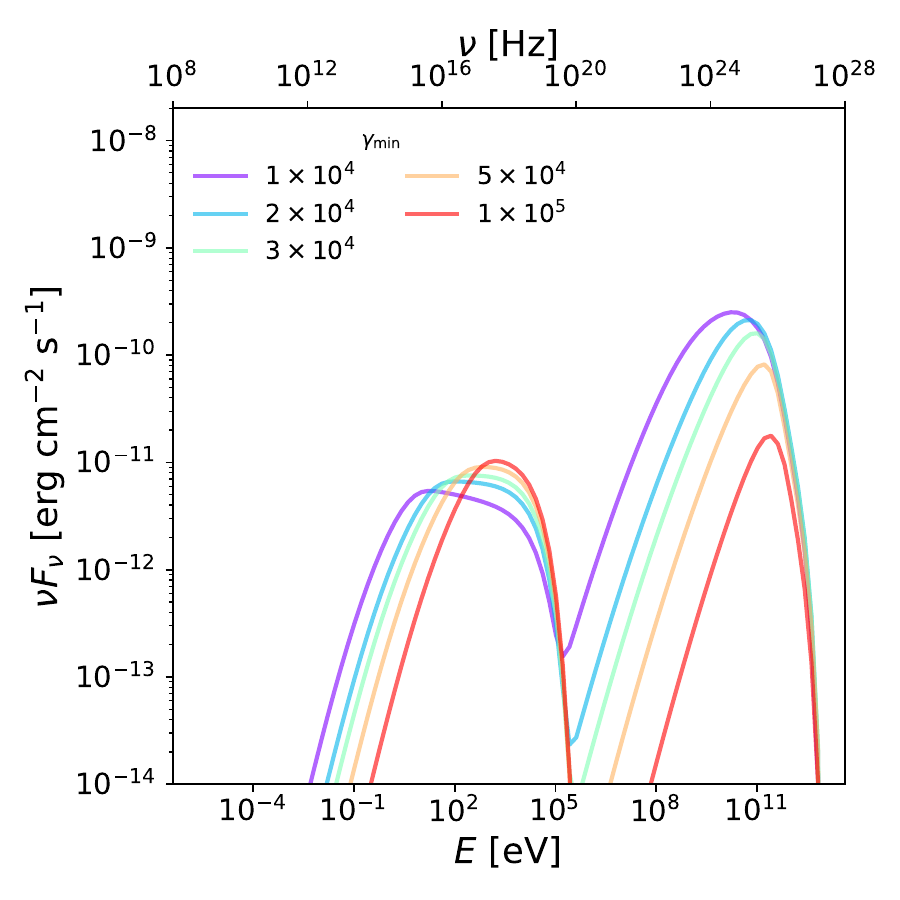}
}\hspace{-5mm}
\quad
\caption{The calculated LCs (left panel) and average SEDs (right panel) of a blob with different $\gamma^{\prime}_{\rm min}$ varying from $1.0\times10^4$ to $1.0\times10^5$. The redshift is assumed as $z=0.116$, and other parameters are set as $r_{\rm blob}=0.01$pc, $\rm \delta_D=20$, $B=0.15$G, $L^{\prime}_{\rm inj}=3.0\times 10^{42} \rm erg\ s^{-1}$, $s=2.8$, $\gamma^{\prime}_{\rm max}=1\times 10^{6}$. From the figure, an increasing $\gamma^{\prime}_{\rm min}$ can result in lower optical and $\gamma$-ray emission accompanyed by an enhancement of the X-ray flux. Here the EBL absorption is taken into account for VHE $\gamma$-ray emission.}
\label{fig:test_gamma_min}
\end{figure*}

\section{Model description}\label{sec:model}

In the stochastic dissipation model, blobs are generated in the jet persistently via random dissipation events, such as magnetic reconnection events \citep[e.g.,][]{2013MNRAS.431..355G, 2015MNRAS.450..183S, 2016MNRAS.462.3325P} or internal shocks \citep[e.g.,][]{2001MNRAS.325.1559S, 2010ApJ...711..445B}. A flare will appear if a strong dissipation event occurs and accelerates an exceptionally large amount of electrons in the accompanying blob(s) compared to others. Therefore, a blazar's emission during a flare consists of the radiation from both the flaring zone and that from non-flaring zones.

Following the treatment in \cite{2023MNRAS.526.5054L}, a conical jet is assumed with a constant half-opening angle $\theta$. Many blobs are distributed inside the jet from a distance $r_0$ to $L+r_0$ away from the central super massive black hole (SMBH). The jet and blobs have a relativistic motion with the bulk Lorentz factor ${\rm \Gamma}=(1-\beta_{\Gamma}^2)^{-1/2}$, where $\beta_{\Gamma} c$ is the jet speed. As a result of relativistic Doppler boosting, the observed emission will be significantly enhanced, characterized by the Doppler factor ${\rm \delta_D}$. In this work, $\theta_{\rm obs} \lesssim 1/\Gamma$ is assumed for blazars leading to $\rm \delta_{D} \approx \Gamma$. For convenience of computation, the jet is divided into segments along the jet axis, labeled by $i$ ($i=1, 2, 3, ..., i_{\rm max}$), from the innermost one at $r=r_0$ to the outermost one at $r=r_0+L$. In each segment, the locally generated blobs share the same physical properties. The radius of blobs in the comoving frame $R^{\prime}$\footnote{Hereafter, quantities in the comoving frame are primed, and quantities in the rest frame are unprimed, unless specified otherwise.}, related to the transverse radius of the segment, $R_{\rm seg}$, through the parameter $\kappa$, such that $R' = \kappa \, R_{\rm seg}$, will be gradually become larger with increasing distance. Assuming conservation of magnetic luminosity along the conical jet \citep[e.g.,][]{1979ApJ...232...34B}, the magnetic field is inversely proportional to the distance. Relativistic electrons are accelerated and injected into blobs over a time period $t^{\prime}_{\rm inj}$, which is considered to be the same as the light crossing timescale of the blob, $R^{\prime}/c$. The injection luminosity of electrons $L^{\prime}_{\rm inj}$ is assumed to be the same in all blobs within a segment, and the injected electron energy distribution (EED) is assumed to be a power-law distribution,
\begin{equation}\label{eq:EED}
Q^{\prime}(r,\gamma^{\prime})=Q^{\prime}_0(r) {\gamma^{\prime}}^{-s}, {\rm for}\ \gamma^{\prime}_{\rm min} < \gamma^{\prime} < \gamma^{\prime}_{\rm max},
\end{equation}
where $\gamma^{\prime}_{\rm min}$ and $\gamma^{\prime}_{\rm max}$ are the minimum and maximum electron Lorentz factors, and $Q^{\prime}_0(r)$ is a normalization constant. The normalization constant $Q^{\prime}_0(r)$ is related to the electron injection luminosity in each blob as $\int Q^{\prime}(r,\gamma^{\prime})\gamma^{\prime} m_{\rm e} c^2d\gamma^{\prime}= L^{\prime}_{\rm inj}$. The temporal evolution of the EED in the blobs can be described by
\begin{equation}\label{eq:continuity}
\begin{split}
\frac{\partial N^{\prime}(\gamma^{\prime}, t^{\prime}, r)}{\partial t^{\prime}}=&-\frac{\partial}{\partial \gamma^{\prime}}[\dot{\gamma^{\prime}}(\gamma^{\prime}, t^{\prime}, r)N(\gamma^{\prime}, t^{\prime}, r)]\\
&-\frac{N^{\prime}(\gamma^{\prime}, t^{\prime}, r)}{t^{\prime}_{\rm esc}(r)}
+Q^{\prime}(\gamma^{\prime}, r),
\end{split}
\end{equation}
where $N^{\prime}(\gamma^{\prime}, t^{\prime}, r)$ is the time-dependent EED, and $t^{\prime}_{\rm esc}(r)=10R^{\prime}/c$ is the escape timescale \citep{2017ApJ...843..109G}. $\dot{\gamma^{\prime}}(\gamma^{\prime}, t^{\prime}, r)=-c \gamma^{\prime}/R^{\prime}-b^{\prime}(\gamma^{\prime}, t^{\prime}, r){\gamma^{\prime}}^2$ is the energy loss rate, where the first term describes adiabatic losses and the second term stands for radiative energy losses caused by synchrotron and IC processes. The Klein-Nishina effect is taken into account. The target soft photons for the IC process have various origins, including the synchrotron photons produced inside the individual blob $u^{\prime}_{\rm blob}(t^{\prime},r)$, the synchrotron photons from the other blobs in the same segment $u^{\prime}_{\rm seg}(t^{\prime},r)$ and the thermal photons from external structures, $u^{\prime}_{\rm ext}(r)$, such as the BLR and the DT. We employ Eqs.~(10) and (11) of \cite{2023MNRAS.526.5054L} to describe the spatial distribution of the energy density of the external radiation as well as their spectra, which were originally proposed by \citet{2007ApJ...660..117C, 2008MNRAS.387.1669G, 2008MNRAS.386..945T, 2012ApJ...754..114H}.

When calculating the radiation of a single blob, the light-travel-time effect is considered \citep[similar to][]{1999MNRAS.306..551C}. Assuming a homogeneous electrons distribution in a blob, the radiation generated by each part of the blob simultaneously will not be detected by the observer at the same time, smoothing the observed LCs. The overall emission of a blob can be written as
\begin{equation}\label{eq:LC blob}
    \begin{split}
    \nu F^{\rm b}_\nu(E,t,r)=&\frac{(2R^{\prime})^{-1}\delta_{\rm D}^4}{4\pi D_L^2}\int_0^{2R^{\prime}} \nu L^{\prime}_\nu\left[E\frac{1+z}{\delta_{\rm D}}, (t-\frac{x}{c})\frac{\delta_{\rm D}}{1+z}, r\right]dx,  \\
    &{\rm with}~L_\nu^{\prime}(t^{\prime}<0)=0.
\end{split}
\end{equation}
With the injection timescale and adiabatic cooling timescale assumed as $R^{\prime}/c$, the active life-time of a blob can be estimated by $t^{\prime}_{\rm blob}=2R^{\prime}/c$ in the co-moving frame, corresponding to $t_{\rm blob}=2R^{\prime} \, (1 + z)/c\delta_{\rm D}$ in the observer's frame.

Another important issue is the $\gamma\gamma$ absorption of VHE photons. The absorption probability per unit pathlength and the total opacity are calculated following the semi-analytical method given by \citet{2009herb.book.....D}. For BL Lacs, the opacity is mainly contributed by two components in the model: one is the radiation generated by the blob itself within the blob of radius $R^{\prime}$, and the other is the radiation emitted from other nearby blobs (or blobs in the same segment) within the segment of size $R^{\prime}/\kappa$. Besides, the external radiation field may also absorb VHE photons although its contribution is usually negligible in HBLs. The opacity for VHE photons will be further addressed in sections below and in the Appendix.

\subsection{Long-term low-state emission}\label{sec:LS}

The long-term activity of a blazar includes a quasi-stable and relatively low flux state, coming from the superposition of numerous blobs along the jet in the model. Following \citet{2023MNRAS.526.5054L}, we define $p(r)\propto r^{-\alpha_{\rm blob}}$ as the probability of a dissipation event occurring per unit time and unit jet length at a distance $r$ in the AGN rest frame. $\alpha_{\rm blob}$ is a model parameter which will influence the resulting SED. Assuming $N$ blobs generated in a period of time $T$, corresponding to a total dissipation rate $\dot{N}$ over the entire jet, the relation between $p(r)$ and the dissipation rate is given by
\begin{equation}\label{eq:generate_rate}
\dot{N}\equiv N/T=\int_{r_0}^{L+r_0} p(r)dr,
\end{equation}
Based on $p(r)$, we can obtain the number of blobs generated in the $i$th segment to be $N_{i,\rm blob}=Tp(r_i)(r_{i+1}-r_i)$ over a period of time $T$.

The observed SEDs of blazars generally suggest a broken power-law distribution of electrons in the emission region(s). The required spectral break is often not consistent with the expectations from the effect of radiative cooling \citep[e.g.][]{1998ApJ...509..608T, 2009MNRAS.397..985G}. With the assumption of a single power-law EED at injection in the model, we further assume that the maximum electron energy (or the Lorentz factor) $\gamma^{\prime}_{\rm max}$ is not the same among all generated blobs. Instead, it follows the distribution
\begin{equation}\label{eq:blob_dis}
    \frac{dN_{i,\rm blob}}{d\gamma^{\prime}_{\rm max}} = C(r_i){\gamma^{\prime}_{\rm max}}^{-\alpha_{\rm \gamma}}, {\rm for}\ \gamma^{\prime}_{\rm max,1}<\gamma^{\prime}_{\rm max}<\gamma^{\prime}_{\rm max,2}
\end{equation}
where $\gamma^{\prime}_{\rm max,1}$ and $\gamma^{\prime}_{\rm max,2}$ are the lower and upper limits of the maximum electron energy, respectively, with $C(r_i)$ being the normalization and $\alpha_{\rm \gamma}$ being a free model parameter. In the $i$th segment, the starting time of activity in the $j$th blob is denoted as $t_{i,j}$. The time-dependent emission of the $i$th segment can be written as
\begin{equation}\label{eq:LC_seg}
\nu F_\nu^{\rm seg,i}(E, t)=\sum_{j=1}^{N_{i,\rm blob}}\nu F_\nu^{\rm b}(E, t, r_i)\Theta(t-t_{i,j}).
\end{equation}
If the duration $T$ is long enough, the average SED of a segment is given by
\begin{equation}\label{eq:SED_seg}
    \overline{\nu F_\nu^{\rm seg,i}}(E)=T^{-1}\int_0^T \nu F_\nu^{\rm seg,i}(E,t)dt.
\end{equation}
The generation and disappearance of blobs can also reach a dynamic equilibrium, and the number of blobs quasi-stably existing in the $i$th segment can be estimated by $N_{i,\rm blob}^{\rm stable}\approx t_{\rm blob}(r_i)p(r_i)(r_{i+1}-r_i)$. For the sake of simplicity, the synchrotron radiation from a blob in the $i$th segment with maximum electron energy $\gamma^{\prime}_{\rm max}$ can be approximately estimated as $\nu L^{\rm b}_{\nu}(r_i, \gamma^{\prime}_{\rm max})\propto Q^{\prime}_0(r_i)\nu^{(3-s)/2}{\rm e}^{-\nu/\nu_{\rm m}}$, where $\nu_{\rm m}={\rm \delta_D}\nu_{\rm B}\gamma^{\prime 2}_{\rm max} / (1 + z)$ is the peak frequency and $\nu_{\rm B}=eB/(2\pi m_e c)$. Therefore, the total synchrotron spectrum can be evaluated as
\begin{equation}\label{eq:alpha_gamma}
    \begin{split}
    \nu L^{\rm seg,i}_{\nu}(r_i)=&\int_{\gamma^{\prime}_{\rm max,1}}^{\gamma^{\prime}_{\rm max,2}} \nu L^{\rm b}_{\nu}(r_i, \gamma^{\prime}_{\rm max})\frac{dN^{\rm stable}_{i,\rm blob}}{d\gamma^{\prime}_{\rm max}}d\gamma^{\prime}_{\rm max},  \\
    &\propto \left\{\begin{aligned}
        \nu^{(4-s-\alpha_{\gamma})/2}\int_{y_2}^{y_1}y^{(\alpha_{\gamma}-3)/2}e^{-y}dy \ {\rm for} \ s>2\\
        \nu^{(2-\alpha_{\gamma})/2}\int_{y_2}^{y_1}y^{(s+\alpha_{\gamma}-5)/2}e^{-y}dy \ {\rm for} \ s<2,
    \end{aligned}
    \right.
\end{split}
\end{equation}
where $y=\nu (1 + z)/({\rm \delta_D}\nu_{\rm B}\gamma^{\prime 2}_{\rm max})$, $y_1=\nu (1 + z) /({\rm \delta_D}\nu_{\rm B}\gamma^{\prime 2}_{\rm max,1})$ and $y_2=\nu (1 + z)/({\rm \delta_D}\nu_{\rm B}\gamma^{\prime 2}_{\rm max,2})$. In the case $s>2$, the integral term is a confluent hypergeometric function $F((\alpha_{\gamma}-1)/2,y_1,y_2)$. For $y_1<1$ i.e. $\nu<{\rm \delta_D}\nu_{\rm B}\gamma^{\prime 2}_{\rm max,1} / (1 + z)$, this term is proportional to $\nu^{(\alpha_{\gamma}-1)/2}$, leading to $\nu L^{\rm seg}_{\nu}(r_i)\propto \nu^{(3-s)/2}$; for $y_2<1<y_1$ i.e. ${\rm \delta_D}\nu_{\rm B}\gamma^{\prime 2}_{\rm max,1} / (1 + z) < \nu < {\rm \delta_D}\nu_{\rm B}\gamma^{\prime 2}_{\rm max,2} / (1 + z)$, this term is a constant, leading to $\nu L^{\rm seg}_{\nu}(r_i)\propto \nu^{(4-s-\alpha_{\gamma})/2}$; and for $y_2>1$ i.e. $\nu > {\rm \delta_D}\nu_{\rm B}\gamma^{\prime 2}_{\rm max,2} / (1 + z)$, it behaves as an exponential cutoff. In general, the spectral shape of a segment can be described as a smoothly connected broken power law with a high-energy cutoff, as shown in Fig.~\ref{fig:test_alpha_gamma}. 

The overall emission of the jet is the superposition of the emission from each segment, i.e.,
\begin{equation}\label{eq:lowstate}
\nu F^{\rm low}_\nu(E, t)=\sum_{i=1}^{i_{\rm max}}\nu F_\nu^{\rm seg,i}(E, t)=\sum_{i=1}^{i_{\rm max}}\sum_{j=1}^{N_i}\nu F_\nu^{\rm b}(E, t, r_i)\Theta(t-t_{i,j}).
\end{equation}
Since the spectral shape of a segment (Eq.~\ref{eq:alpha_gamma}) does not depend on the distance, the superimposed spectrum will also be described by a broken power-law, which is consistent with the observed SEDs of blazars.
Note that in the radio band, where synchrotron self-absorption is important, the overall spectral shape is demonstrated to be $F_\nu^{\rm low} \propto \nu^{2-\alpha_{\rm blob}}$ \citep{2023MNRAS.526.5054L}. The flat radio spectrum as usually observed from blazars is obtained for $\alpha_{\rm blob}=2$.

While modeling the low-state emission, the variation of flux in the X-ray and the $\gamma$-ray band is negligible compared to the huge enhancement during the flare. For simplicity, we neglect its time dependence and take the long-term average flux as the low-state flux, which can be obtained by 
\begin{equation}\label{eq:average}
    \overline{\nu F^{\rm low}_\nu}(E)=T^{-1}\int_0^T \nu F^{\rm low}_\nu(E,t)dt.
\end{equation}

\subsection{Short-term flare emission}\label{sec:flare}

Flares are intense activity periods of blazars, usually with short duration and considerable flux variation amplitude. Though the origin of blazar flares is not clearly understood yet, it is presumably attributed to strong and rapid energy dissipation in a relatively compact region of the jet. In our model, this compact and energetic flaring zone (blob) can be generated anywhere inside the jet, mainly determined by the observed variability timescale. The distance of the flaring zone from the SMBH determines its ambient electromagentic environment. When the distance from the SMBH satisfying $r_{\rm flare}<r_0$, there is no other blobs around the flaring blob, and hence no synchrotron photons from other blobs to seed IC radiation. On the contrary, for $r_{\rm flare}>r_0$, the synchrotron photons from other blobs may enhance the IC radiation and contribute to the $\gamma\gamma$ absorption opacity. The radiation of the flaring blob follows the same form as shown in Eq.~(\ref{eq:LC blob}). When modeling a specific outburst event composed of several flares, we introduce a number of $n_{\rm max}$ flaring blobs to be triggered at $t^{\rm flare}_n, n=1,2,...n_{\rm max}$, consecutively. The calculated LC can be written as
\begin{equation}\label{eq:LC_flare}
\nu F_\nu^{\rm flare}(E, t)=\sum_{n=1}^{n_{\rm max}}\nu F_\nu^{\rm b}(E, t, r_{n})\Theta(t-t^{\rm flare}_n),
\end{equation}
where $r_{n}$ represents the location of the $n$th flaring blob. The average SED of a selected period of time $T_1$ to $T_2$ can be given by
\begin{equation}\label{eq:SED_flare}
    \overline{\nu F_\nu^{\rm flare}}(E)=(T_2-T_1)^{-1}\int_{T_1}^{T_2} \nu F_\nu^{\rm flare}(E,t)dt.
\end{equation}
The radiation during the outburst is the superposition of the low-state and flaring components,
\begin{equation}\label{eq:emission_total}
\nu F_\nu^{\rm total}(E, t)=\nu F_\nu^{\rm low}(E, t)+\nu F_\nu^{\rm flare}(E, t)\approx \overline{\nu F_\nu^{\rm low}}(E)+\nu F_\nu^{\rm flare}(E, t).
\end{equation}

Modeling of the emission of a flaring blob is similar to that in the classical one-zone model. Main parameters of each flaring blob include $\rm \delta_D$, $r$, $B$, $L^{\prime}_{\rm inj}$, $s$, $\gamma^{\prime}_{\rm min}$ and $\gamma^{\prime}_{\rm max}$. While flares can be produced by increasing the Doppler factor of the blob as the predicted flux depends on $\delta_{\rm D}^4$, the value of $\delta_{\rm D}$ will be kept the same as those in the low state for the purpose of this study. The distance of the blobs can be determined based on the observed variability timescale of identified flares as long as $\rm \delta_D$ is fixed. $B$ and $L^{\prime}_{\rm inj}$ can be estimated based on X-ray and $\gamma$-ray flux, while the three EED-related parameters can be estimated from the observed spectrum. A parameter seldom discussed in the classical one-zone model is the minimum electron energy $\gamma^{\prime}_{\rm min}$, which is usually fixed to $10-100$, a value sufficiently small to account for the infrared-optical emission of the blazar. In our model, $\gamma^{\prime}_{\rm min}$ of a flaring blob may in principle be much larger than this value, because the infrared-optical emission of the blazar could be attributed to the low-state emission. The influence of $\gamma^{\prime}_{\rm min}$ on the resultant LCs and SED is illustrated in Fig.~\ref{fig:test_gamma_min}. Here, for a more straightforward comparison, the amount of the high-energy electrons is kept constant (i.e., the normalization of EED is kept the same) to show the influence caused by the absence/presence of low-energy electrons. The low-frequency (i.e., optical) flux increases by setting a small $\gamma_{\rm min}'$, because the typical Lorentz factor of electrons radiating at the optical band via the synchrotron process is $\sim 10^4$ for the given magnetic field. On the other hand, however, a small $\gamma_{\rm min}'$ decreases the X-ray flux, whereas the injection rate of electrons that can emit in the X-ray band remains the same for different cases of $\gamma_{\rm min}'$. This is because the increased optical radiation at small $\gamma_{\rm min}'$ case provides additional targets for the IC scattering and hence suppresses the synchrotron radiation of those high-energy electrons. For a similar reason, the IC flux in the GeV-TeV band increases for a decreasing $\gamma_{\rm min}'$, especially considering that the presence of the additional optical radiation field causes the IC scattering to proceed in the Thomson regime. 

\section{Application}\label{sec:app}

In this section, the stochastic dissipation model is applied to explain both the low state and two remarkable outbursts of PKS~2155-304 observed in 2006. 

\begin{figure*}
\centering
\subfigure{
\includegraphics[width=0.9\columnwidth]{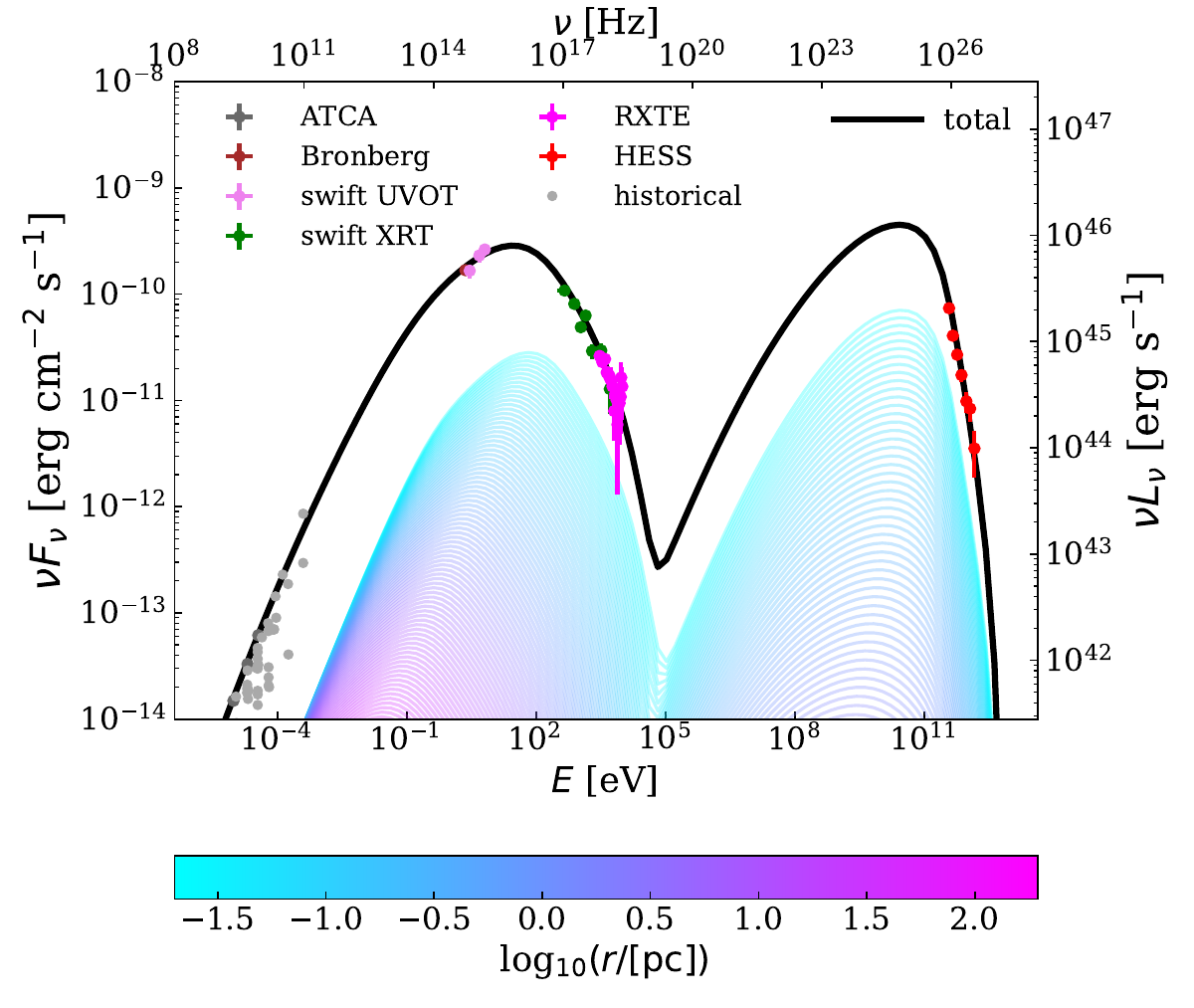}
}\hspace{-5mm}
\quad
\subfigure{
\includegraphics[width=0.9\columnwidth]{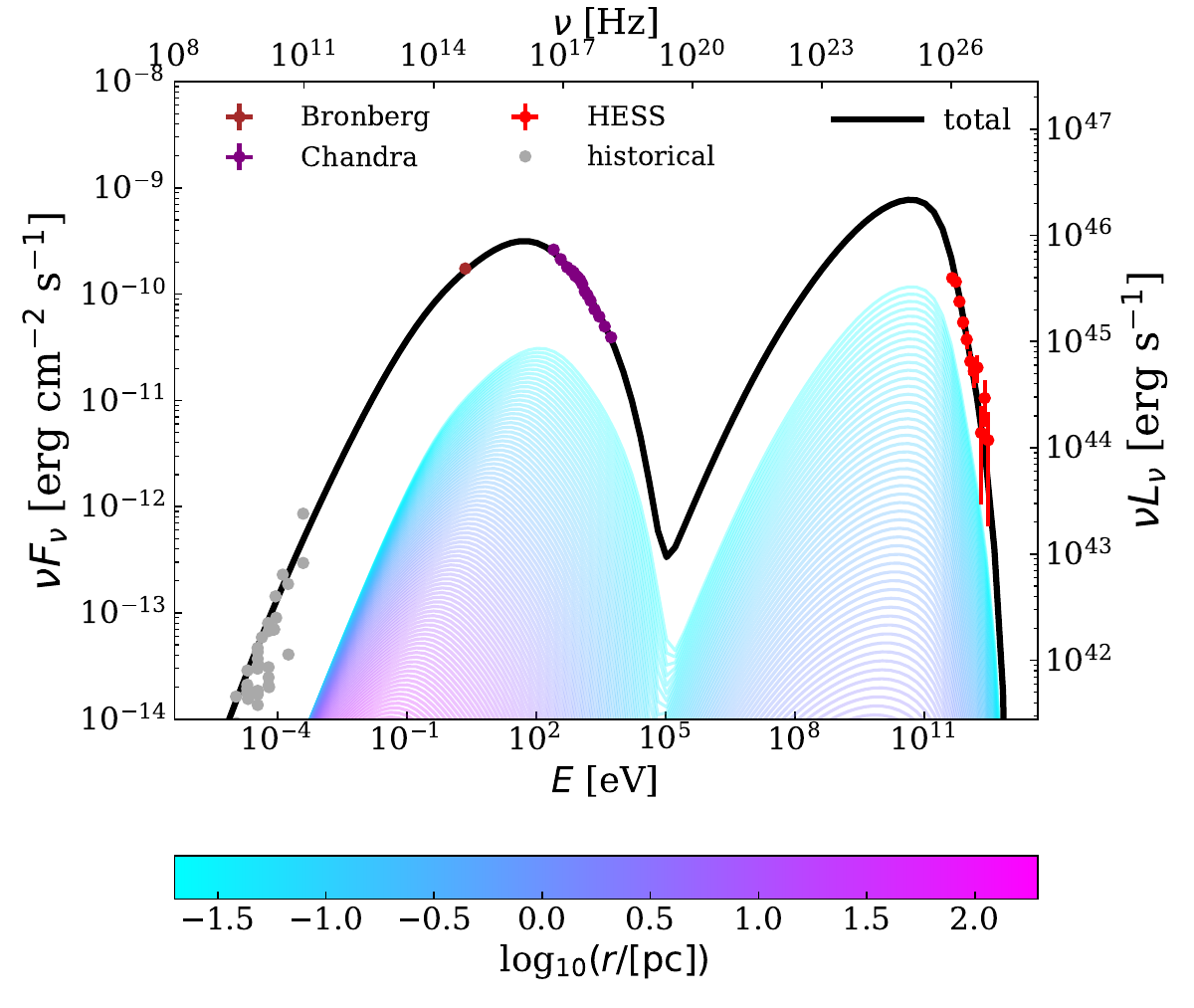}
}\hspace{-5mm}
\quad
\caption{The average SEDs calculated for the two low states. The left panel corresponds to LS1, while the right panel represents LS2. The colored dots represent the observations in different energy bands as labeled in the legend. The black solid line is the overall emission generated by the jet, while the gradient color lines show the individual SEDs from each segment at different distances.}
\label{fig:LS_SED}
\end{figure*}

\begin{table*}
\begin{minipage}[t][]{\textwidth}
\caption{Summary of low states Parameters.}
\label{tab:LS}
\centering
\begin{tabular}{cccc}
\hline\hline
Free parameters	&	LS1  &   LS2	&	Notes	\\
\hline	
$z$ & 0.116 & 0.116 & Redshift \\
$D_{\rm L}$[Mpc]   & 483.5 & 483.5 &  Luminosity distance \\
$\delta_{\rm D}$	&	20	&	20 &	Doppler factor \\
$L_{\rm D}$ [erg/s]	&	$1.0\times10^{42}$		&	$1.0\times10^{42}$	&	Disk luminosity	\\
$r_0$ [pc]	&	$2.0\times10^{-2}$	&	$2.0\times10^{-2}$	&	 Position of jet base	\\
$L$ [pc]	&	200 &	200 &	Jet length	\\
$\theta~[\circ]$	&	5	&	5	&	Jet's half-opening angle	\\
$\kappa$	&	0.3	&	0.3 &	 Ratio of blob's radius to its segment's radius 	\\
$\dot{N}~\rm [s^{-1}]$	&	$0.93$		&	$0.93$  &	Blob generation rate of the entire jet	\\
$\alpha_{\rm blob}$	&	2.0	&	2.0	&	 Index of the dissipation probability $p(r)$	\\
$B^{\prime}(r_0)$ [G]	&	0.107	 &  0.080 &	Magnetic field at jet base	\\
$L^{\prime}_{\rm inj}$~[erg/s]	&	$7.3\times10^{39}$		&	$1.0\times10^{40}$	&	Injection electron luminosity at jet base	\\
$s$	&	2.0	&	2.0	&	The spectral index of electron energy distribution	\\
$\gamma^{\prime}_{\rm min}$	&	4.0$\times10^{3}$	&	5.0$\times10^{3}$	&	Minimum electron Lorentz factor	\\
$\gamma^{\prime}_{\rm max,1}$	&	7.0$\times10^{4}$		&	1.0$\times10^{5}$	&	Lower limit of maximum electron Lorentz factor\\
$\gamma^{\prime}_{\rm max,2}$	&	9.0$\times10^{5}$	&	1.2$\times10^{6}$		&	Upper limit of maximum electron Lorentz factor\\
$\alpha_{\rm \gamma}$   & 2.8   & 2.5 & $N_{\rm blob} \propto {\gamma^{\prime}_{\rm max}}^{-\alpha_{\rm \gamma}}$ \\
\hline						
\end{tabular}
\end{minipage}
\end{table*}

\begin{figure*}
\centering
\subfigure{
\includegraphics[width=0.9\columnwidth]{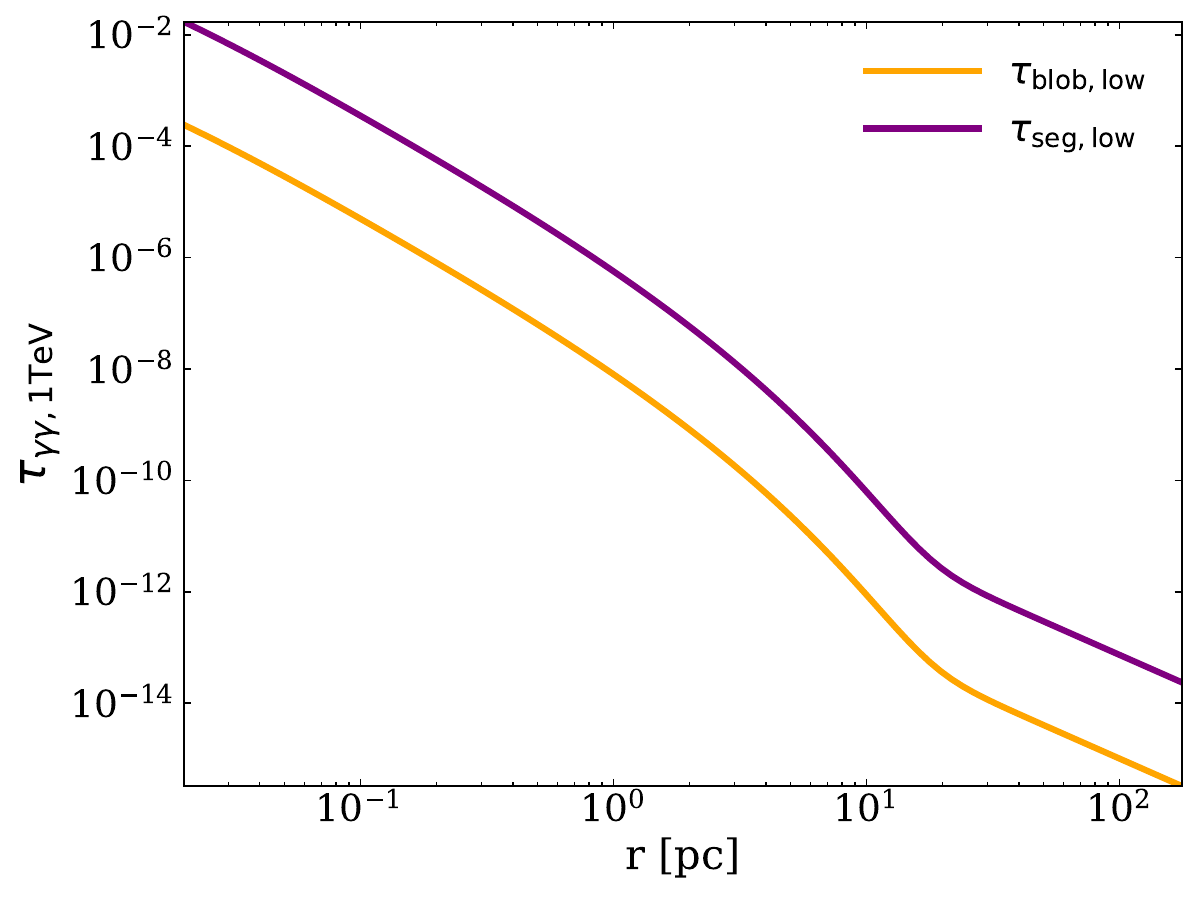}
}\hspace{-5mm}
\quad
\subfigure{
\includegraphics[width=0.9\columnwidth]{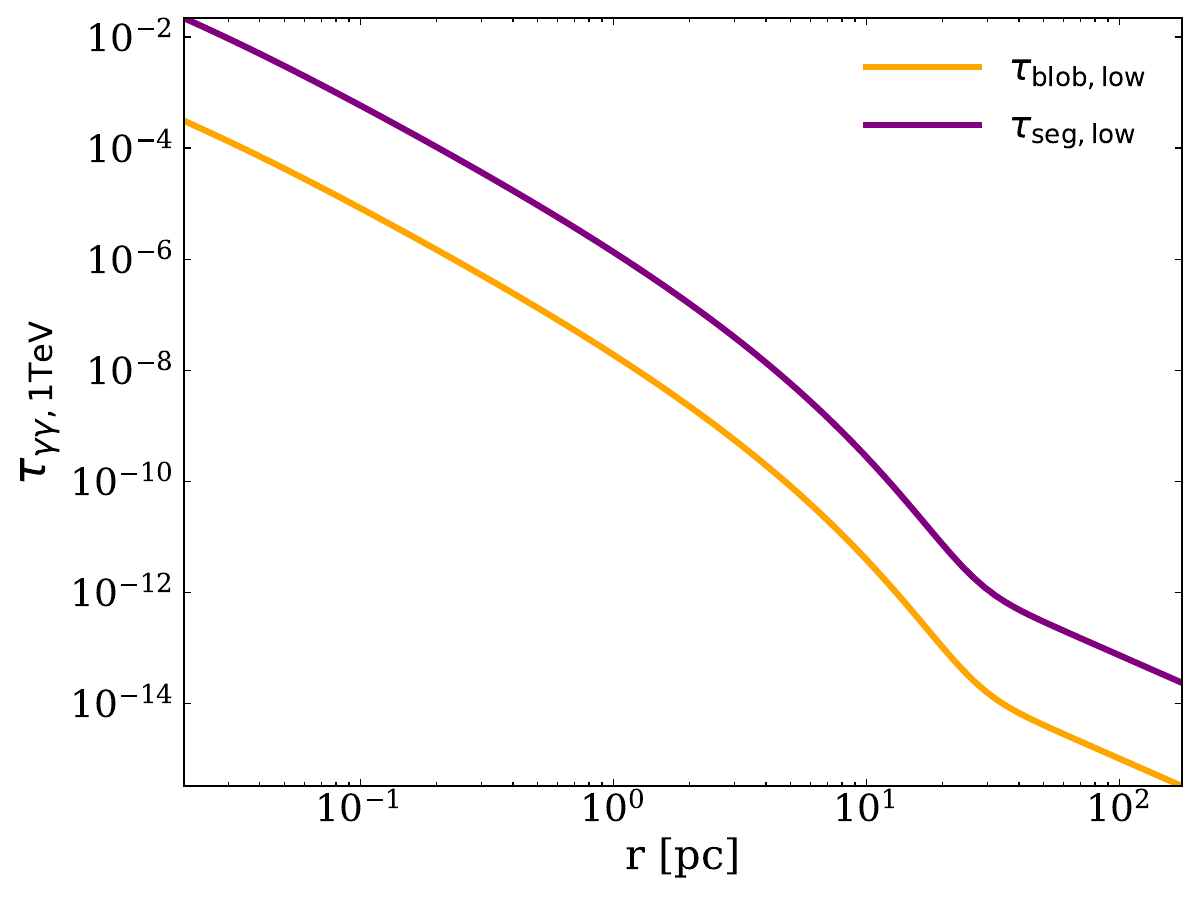}
}\hspace{-5mm}
\quad
\caption{Opacity of $\gamma\gamma$ absorption during the low state for the first outburst (left panel) and the second outburst (right panel). The orange solid line represents the opacity of photons at 1\,TeV induced by the blob’s synchrotron photons, while the purple solid line is the opacity of photons at 1\,TeV caused by segment’s synchrotron photons.
\label{fig:LS tau}}
\end{figure*}

\subsection{Modeling low states}

Observations of PKS~2155-304 by H.E.S.S. in July 2006 detected two extraordinary outbursts. The first outburst took place in the early hours of MJD~53944. This outburst lasted for more than one hour and gradually decayed to a low state. The low state remained quasi-stable until midnight of MJD~53944. This quasi-stable period is chosen as the reference low state of the first outburst and is denoted as LS1. After the first remarkable outburst, multiwavelength observations were devoted to this blazar. The average SED during the night of MJD~53944 (LS1) is shown in the left panel of Fig.~\ref{fig:LS_SED}. Later, another giant outburst occurred 44 hours after the first one. The gamma-ray flux of the blazar was already in the flaring state at the beginning of observation time of that night, which was around MJD~53945.85. After 4 -- 5 hours of observations, the gamma-ray flux underwent a relatively slow decay process in the early hours of MJD~53946. In the period from MJD~53946.10 to MJD~53946.15, the emission reached the lowest level and was comparable to the observed flux level during the night of MJD~53946. Hence, the period between MJD~53946.10 and MJD~53946.15 is considered to be the low state of the second outburst and is denoted as LS2. The average SED of this period is shown in right panel of Fig.~\ref{fig:LS_SED}.

In general, the measured SEDs of LS1 and LS2 are similar. In LS2, the X-ray and $\gamma$-ray fluxes are a bit higher, with a harder spectrum. The hardening in the $\gamma$-ray band is insignificant due to extragalactic background light (EBL) absorption, while spectral index in the X-ray band changes more obviously ($\Gamma_{\rm X,LS1}\approx2.84\pm0.09, \Gamma_{\rm \gamma,LS1}\approx3.99\pm0.12$ and $\Gamma_{\rm X,LS2}\approx2.65\pm0.07, \Gamma_{\rm \gamma,LS2}\approx3.67\pm0.18$). 
This may be achieved with a smaller $\alpha_{\gamma}$ in LS2, which leads to a harder X-ray spectrum as discussed in Section~\ref{sec:LS}. 
The model-predicted SEDs in LS1 and LS2 are displayed in Fig.~\ref{fig:LS_SED} and the parameters are listed in Table~\ref{tab:LS}. The Doppler factor of the blazar jet is fixed to be 20 in both periods. As the interval time between these two periods is relatively small and the SED difference is not large, we try to keep as many parameters unchanged as possible, leaving only a few of them independent. 

With these sets of parameters, the IC emission is found to be dominated by target photons from other blobs in the same segment. The reason is as follows. The energy density of the synchrotron photons in the selected blob itself can be estimated as $u^{\prime}_{\rm syn,blob}=L^{\prime}_{\rm syn}/(4\pi {R^{\prime}}^2 c)$, and the energy density of synchrotron photons from other blobs can be written as $u^{\prime}_{\rm syn,seg}\approx N_{i,\rm blob}^{\rm stable}L^{\prime}_{\rm syn}/(S_{\rm seg} c)$, where in the $i$th segment $S_{\rm seg,i}=\pi(r^2_i+r^2_{i+1})\tan^2\theta+\pi(r^2_{i+1}-r^2_i\tan\theta/\cos\theta)$. 
In the case $\alpha_{\rm blob}=2$, the number of stably existing blobs is almost the same in each segment $N_{i,\rm blob}^{\rm stable}\approx N_{\rm blob}^{\rm stable}$, and will only be affected by the total dissipation rate. The ratio $u^{\prime}_{\rm syn,seg}/u^{\prime}_{\rm syn,blob}\sim N_{\rm blob}^{\rm stable}\kappa^2$ determines the relative importance of these two components. Due to the same values of $\dot{N}$ and $\kappa$ adopted for LS1 and LS2, the ratios in both these period are about 40, implying a dominant synchrotron photon density from nearby blobs. 

The dominance of the radiation density from nearby blobs is also reflected in the $\gamma\gamma$ absorption opacity. As shown in Fig.~\ref{fig:LS tau}, the orange solid line represents the opacity of photons at 1 TeV induced by the blob's synchrotron photons, while the purple solid line is the opacity of photons at 1 TeV caused by the segment's synchrotron photons. The ratio of the former to the latter reaches $\tau^{\rm seg}_{\gamma\gamma}/\tau^{\rm blob}_{\gamma\gamma} \sim \kappa^{-1}u^{\prime}_{\rm syn,seg}/u^{\prime}_{\rm syn,blob}\sim 100$. We can also see from the figure that the $\gamma\gamma$ absorption effect is weak so that TeV photons are able to escape from the jet easily. 

\begin{figure*}
\centering
\subfigure{
\includegraphics[width=1.0\columnwidth]{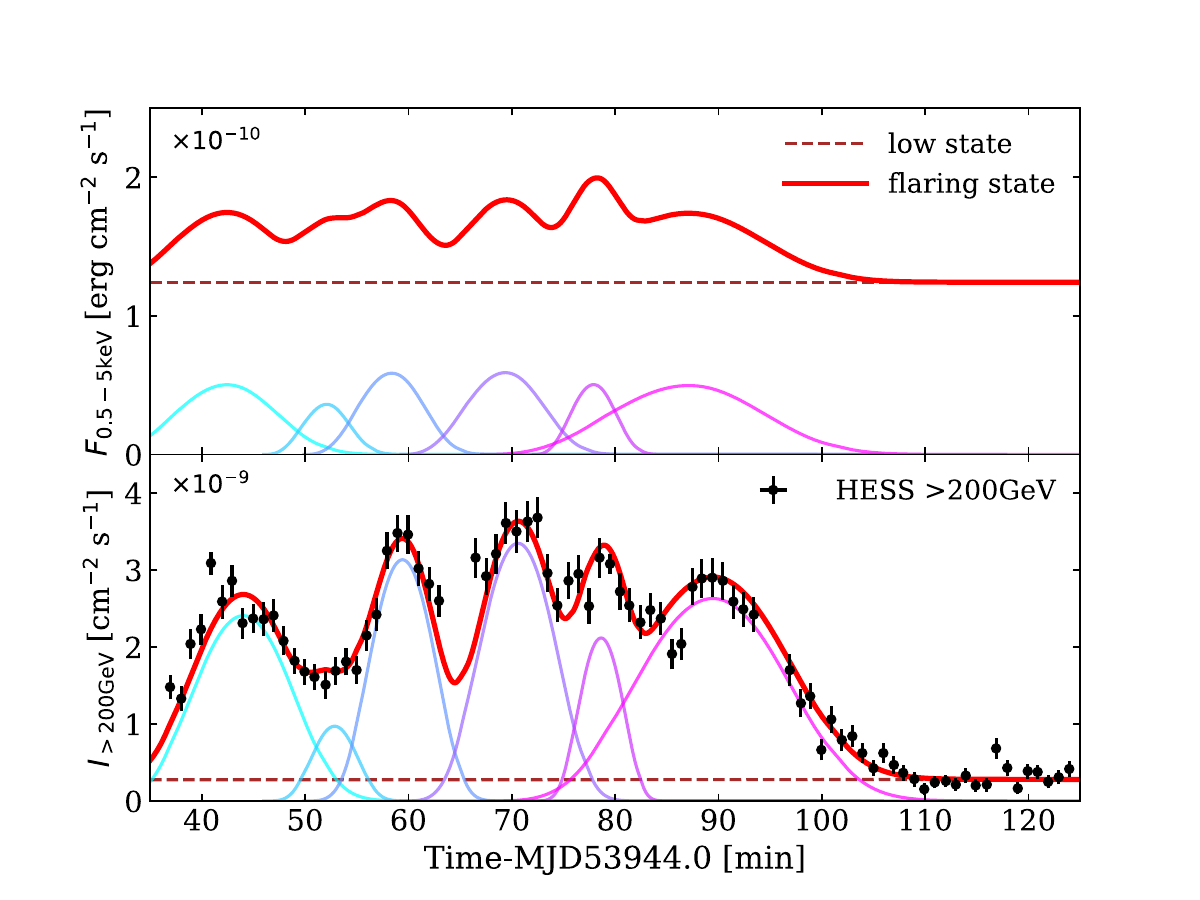}
}\hspace{-5mm}
\quad
\subfigure{
\includegraphics[width=1.0\columnwidth]{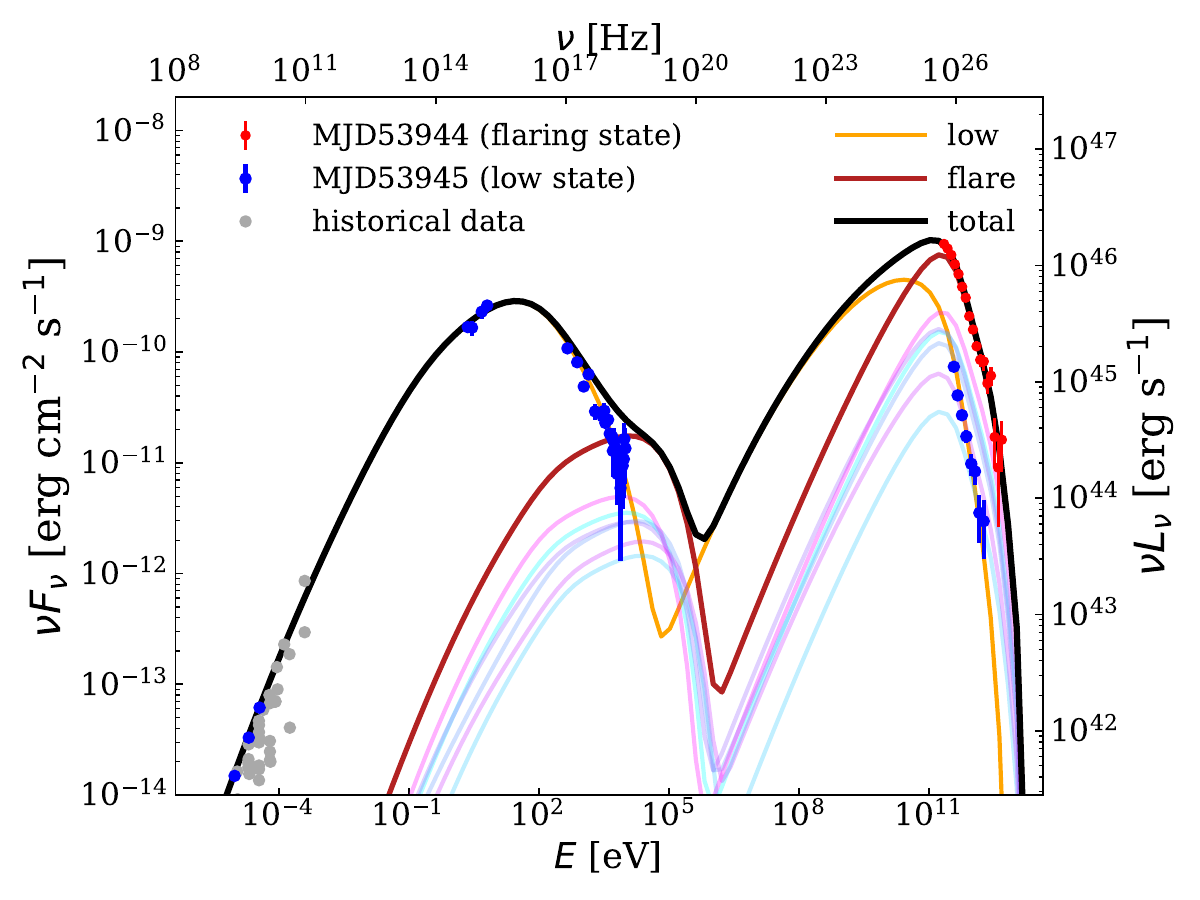}
}\hspace{-5mm}
\quad
\caption{Simulated LCs (left panel) in the X-ray and $\gamma$-ray bands and the average SED (right panel) during the outburst on MJD~53944. In the left panel, the brown dashed line represents the average flux of LS1, the red solid line shows the total flux during the flare and solid lines with gradient colors from cyan to purple are fluxes generated by different blobs with strong injection. In the right panel, the orange solid line represents the SED of LS1, the crimson solid line is the average flux of flaring blobs during the outburst, and the black solid line is the sum of the two components (i.e., the total average flux during the first outburst). The lines with varied colors are the average contribution of each blob.}
\label{fig:MJD53944}
\end{figure*}

\begin{figure*}
\centering
\subfigure{
\includegraphics[width=0.68\columnwidth]{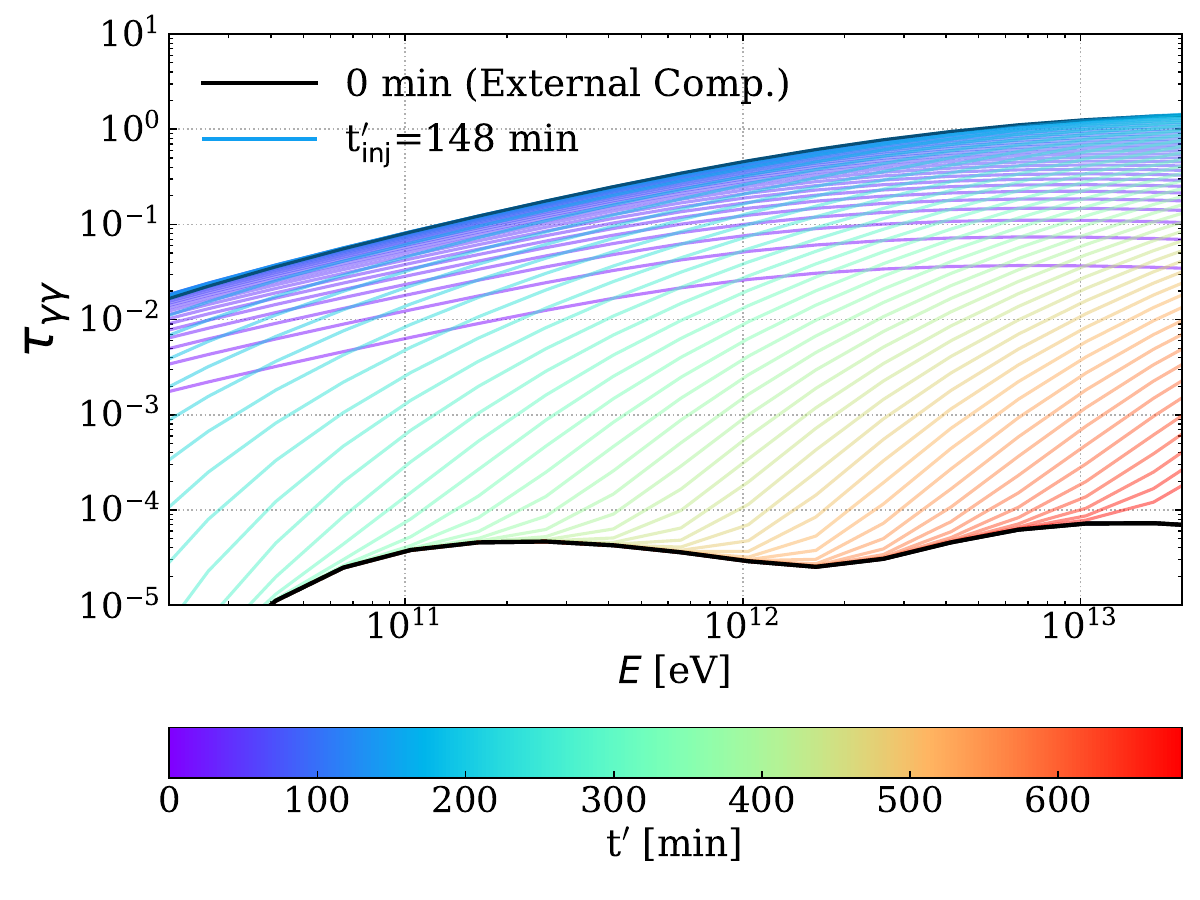}
}\hspace{-5mm}
\quad
\subfigure{
\includegraphics[width=0.68\columnwidth]{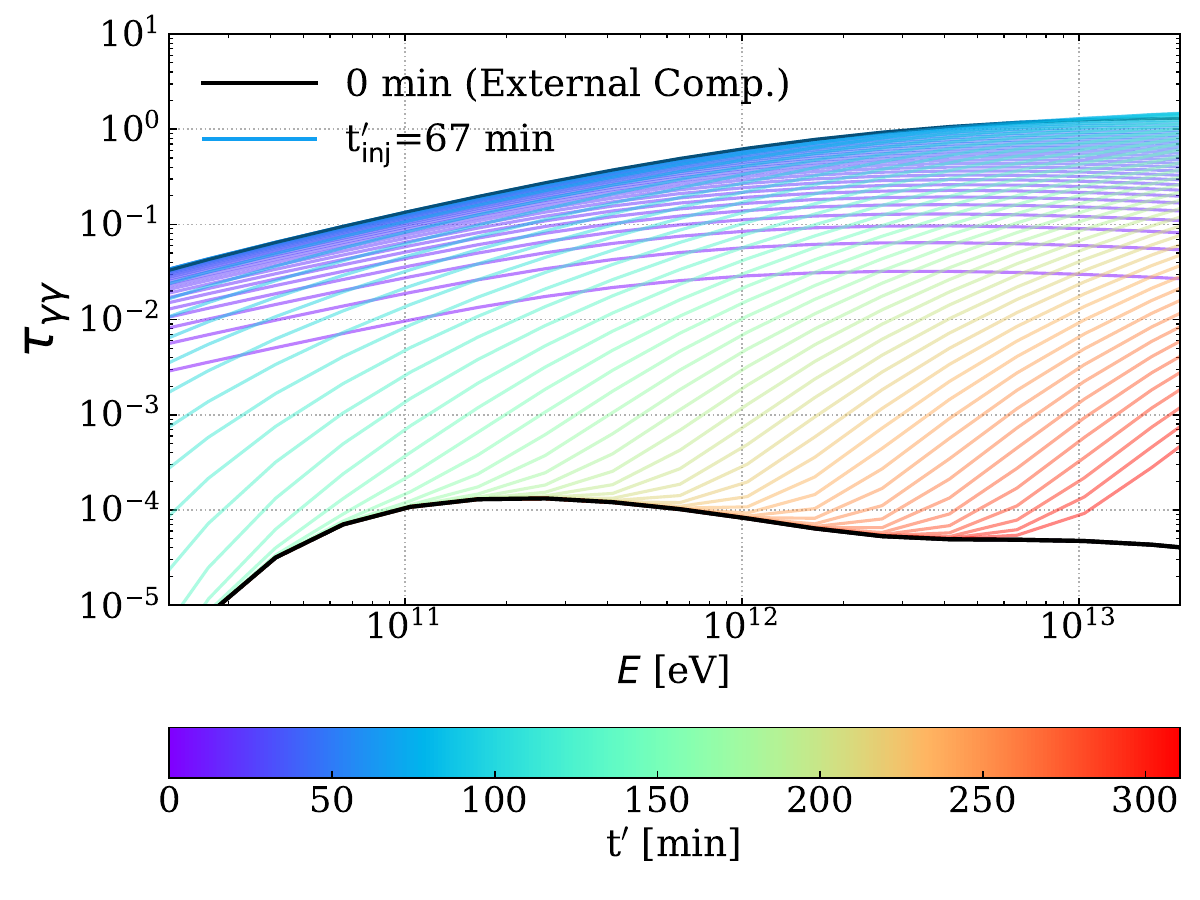}
}\hspace{-5mm}
\quad
\subfigure{
\includegraphics[width=0.68\columnwidth]{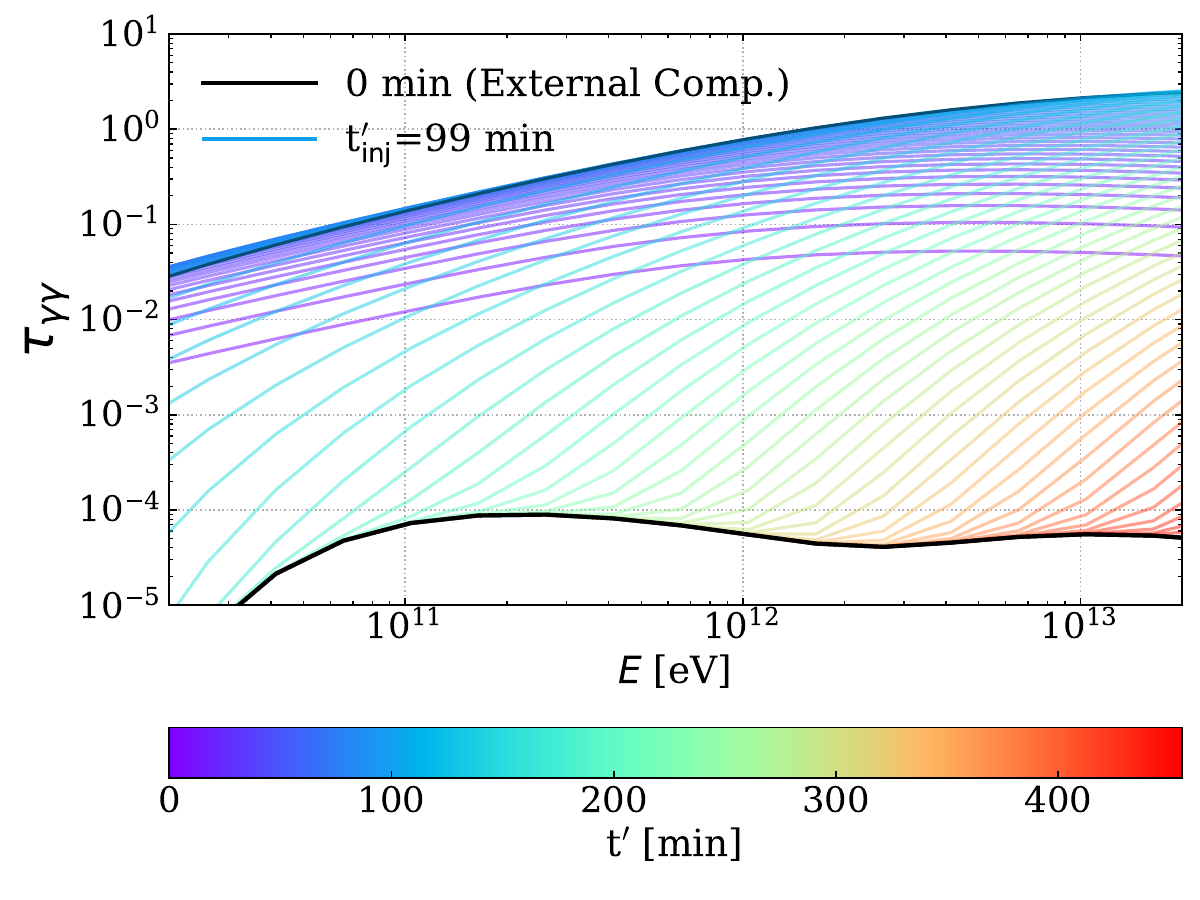}
}\hspace{-5mm}
\quad
\subfigure{
\includegraphics[width=0.68\columnwidth]{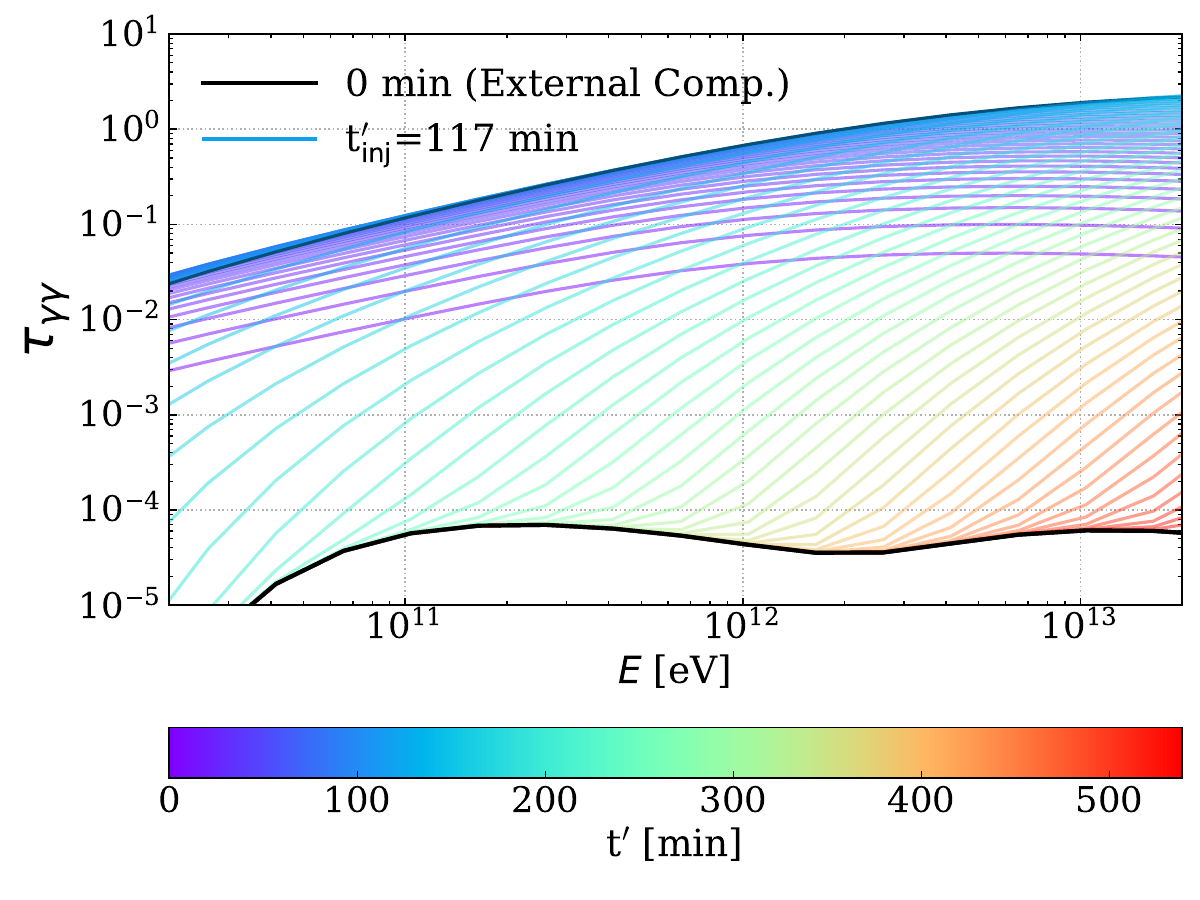}
}\hspace{-5mm}
\quad
\subfigure{
\includegraphics[width=0.68\columnwidth]{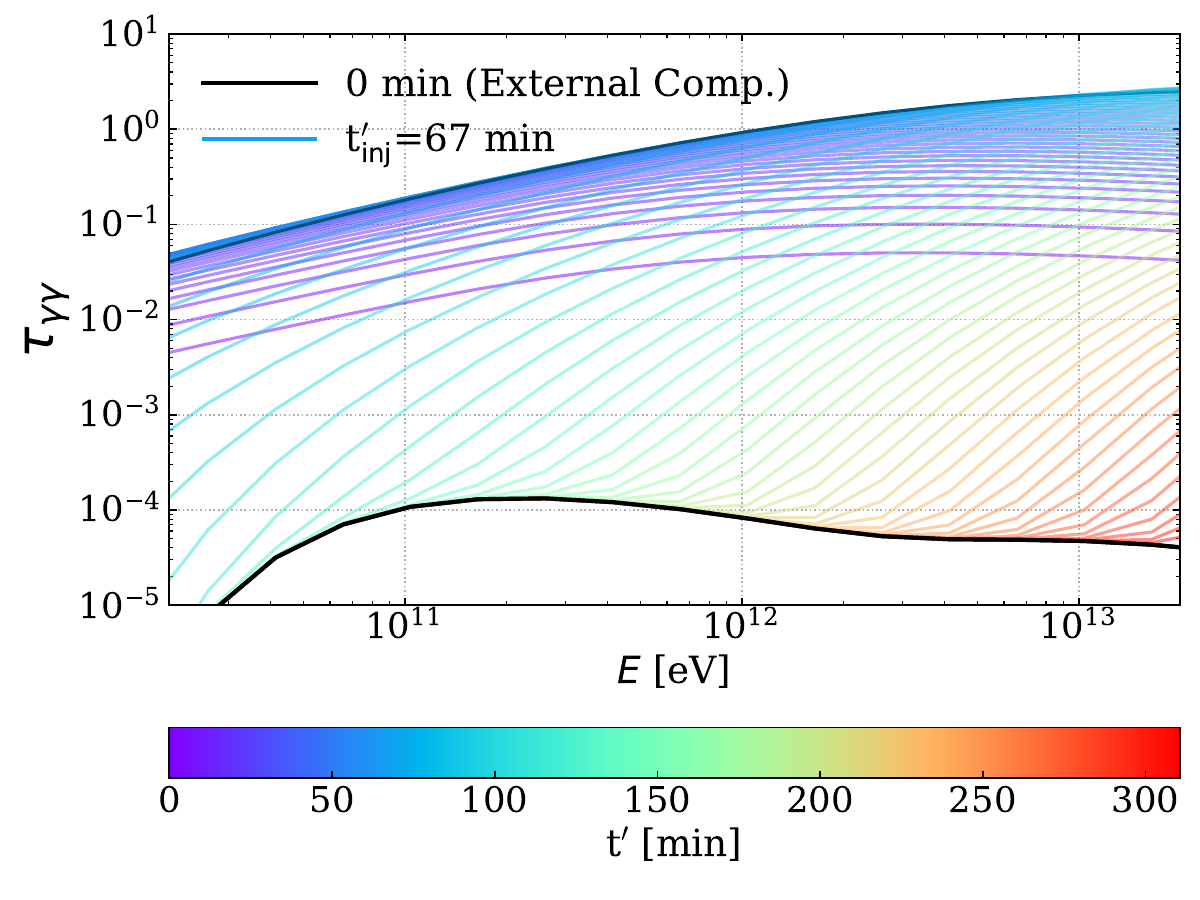}
}\hspace{-5mm}
\quad
\subfigure{
\includegraphics[width=0.68\columnwidth]{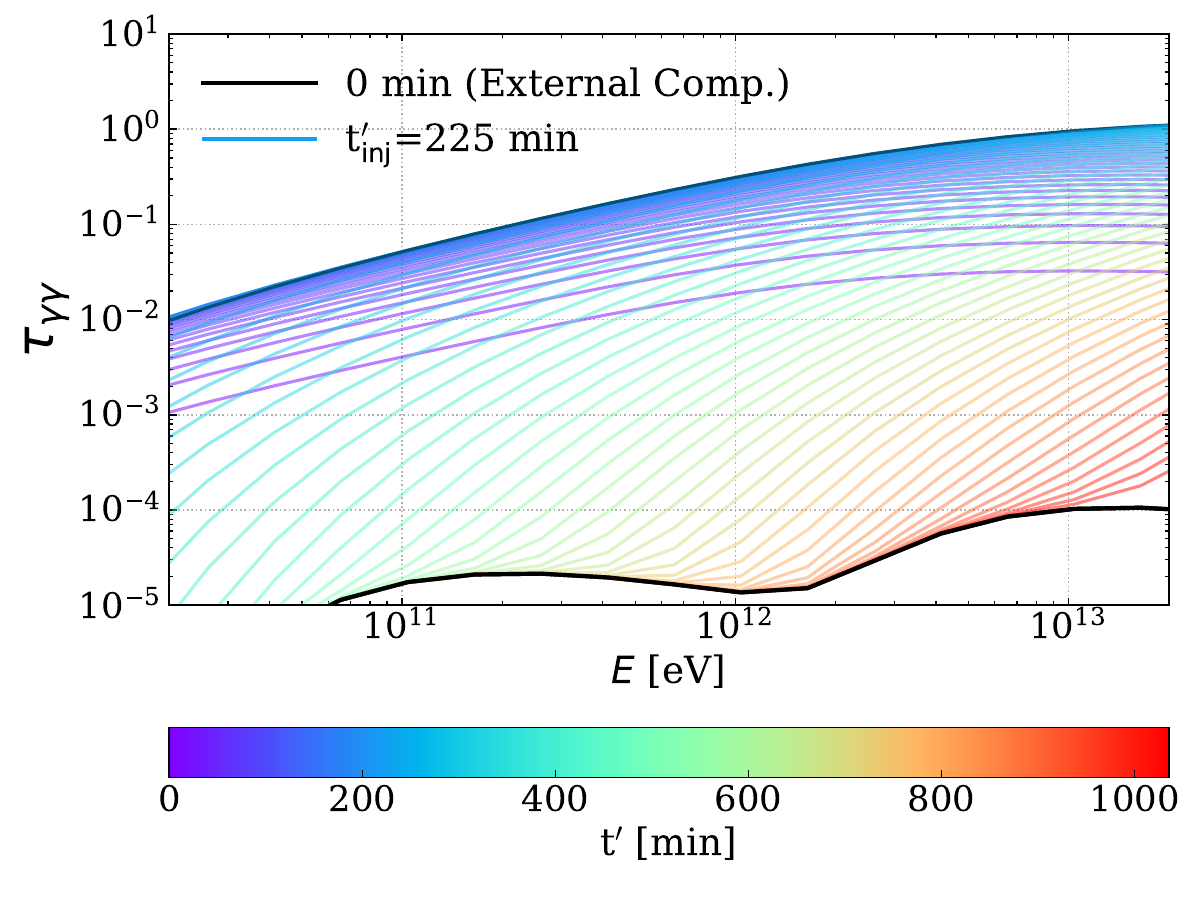}
}
\caption{Energy-dependent opacity of $\gamma\gamma$ absorption of each blob during the outburst on MJD~53944. The black solid line is the opacity at the starting time attributed to stable soft photons. The gradient color lines represent the opacity calculated at different moments from 0 min to $5t^{\prime}_{\rm inj}$. The injection timescales are labeled at the left upper corner of each figure.
\label{fig:MJD53944 tau}}
\end{figure*}

\begin{figure}
\centering
\subfigure{
\includegraphics[width=0.9\columnwidth]{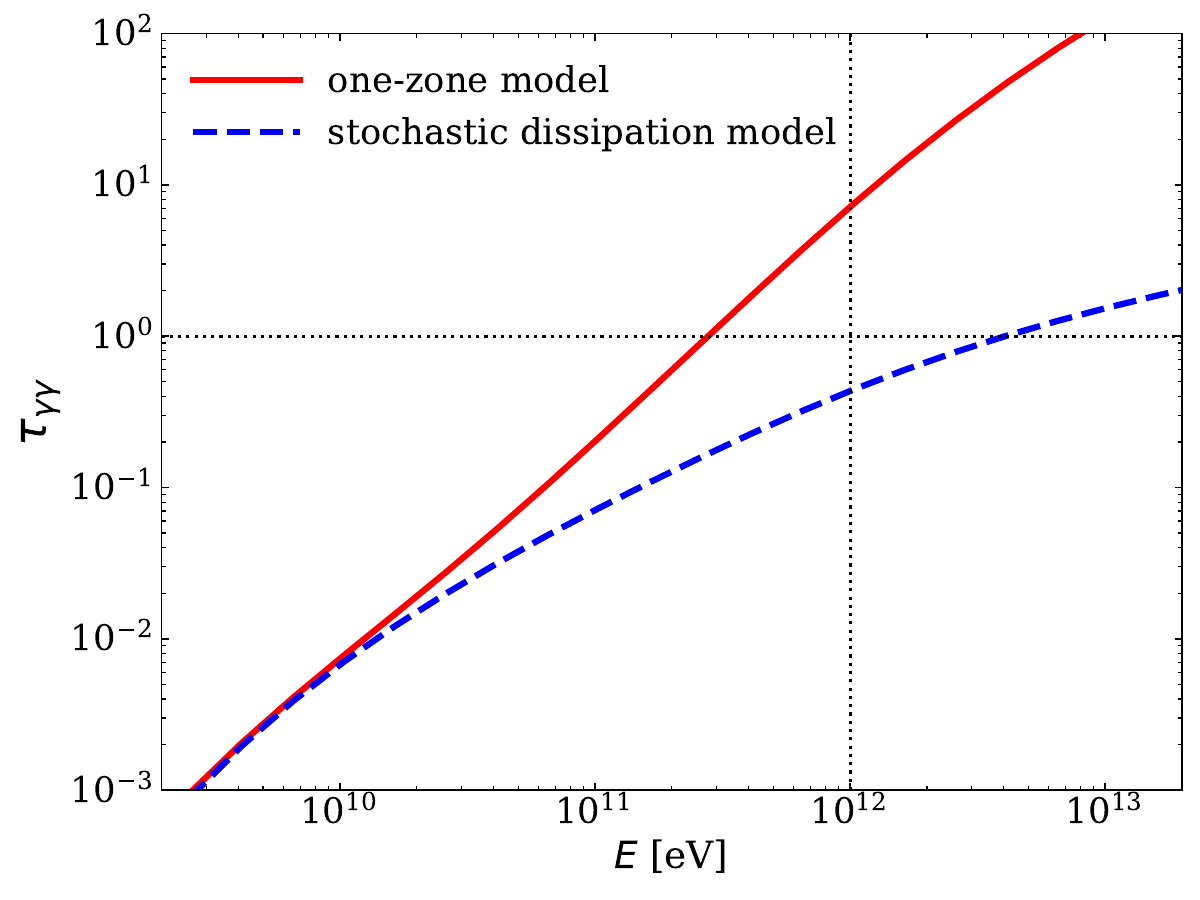}
}\hspace{-5mm}
\quad
\caption{Energy-dependent opacity to $\gamma\gamma$ absorption for flare 5 of the outburst on MJD~53944 for a one-zone model (red solid line) 
and the stochastic dissipation model (blue dashed line). Based on the observed variability timescale in the first flare, an emission region with $\delta_{\rm D}=20$ and $R^{\prime}=1.21\times 10^{14}$\,cm is adopted for the one-zone model. 
\label{fig:Comparison}}
\end{figure}

\subsection{Modeling the outburst on MJD~53944}

To reproduce the LC and average SED of PKS~2155-304 in the night of MJD~53944, six strong dissipation events are assumed to take place, each corresponding to an identified flare. In Fig.~\ref{fig:MJD53944}, the lower-left panel shows the $\gamma$-ray LC measured by H.E.S.S. above 200\,GeV and the fitting result, whereas the corresponding LC predicted in the X-ray band from 0.5\,keV to 5\,keV is shown in the upper-left panel. The right panel of Fig.~\ref{fig:MJD53944} showcases the average SED during the outburst. Note that there is no simultaneous X-ray observation during this period, and hence parameters cannot be well constrained. To avoid tuning too many parameters, we artificially impose some constraints on parameters, noting that these constraints do not necessarily have to be fulfilled. 
The Doppler factor is kept the same as that in the low state; the injected EED is the same for all six flares; the distance of the blobs appearing is determined by the variability timescale once the Doppler factor is fixed; the magnetic field of each small flare is imposed to be inversely proportional to the distance. Parameters for all six flares are listed in Table.~\ref{tab:MJD53944}.

The temporal evolution of the $\gamma$-ray opacity of each flare is shown in Fig.~\ref{fig:MJD53944 tau}. Different colors of curves represent $\tau_{\gamma\gamma}(E_{\gamma})$ at different evolutionary times ranging from 0 to $5t^{\prime}_{\rm inj}$ since the initial particle injection. 
At the beginning time, the $\gamma\gamma$ absorption is mainly due those quasi-stable radiation component such as the radiations of BLR and DT, and from those non-flaring blobs in the jet (i.e., the low-state component), as shown with the black solid curve. The opacity increases as the electron injection is being carried on, because the soft radiation field builds up by the synchrotron radiation of injected electrons. When $t^{\prime}=t^{\prime}_{\rm inj}$, $\tau_{\gamma\gamma}(E_{\gamma})$ achieves the largest value. After that, the electron injection ceases and the synchrotron radiation field fades away. The opacity therefore decreases with time and finally returns to the level of the beginning time. We see that the highest opacity for 1\,TeV photons in all the flares during the evolution is smaller than unity. Among these flares, flare 5 has the shortest variability timescale, which is observed to be about 200\,s. In Fig.~\ref{fig:Comparison}, we compare the expected average $\gamma$-ray opacity of this flare between a classical one-zone model and the stochastic dissipation model, fixing $\delta_{\rm D}=20$ for the one-zone model. To reproduce the 200\,s variability timescale, a compact blob with a radius of $R^{\prime}=1.21\times 10^{14}$\,cm is needed. In the one-zone model, since all the emission of the blazar needs be produced in this compact region, the resulting $\gamma\gamma$ absorption opacity exceeds unity above a few hundred GeV, as shown with the red solid line. In the multi-blob framework, the blazar's emission is shared among many blobs, and the synchrotron flux from the flaring blob itself is just a small fraction of the total observed one. As a result, the expected $\tau_{\gamma\gamma}$, indicated by the blue dashed line, is smaller than that in the one-zone model by $1-2$ orders of magnitude above a few hundred GeV.

\begin{table*}
\begin{minipage}[t][]{\textwidth}
\caption{Summary of Parameters of Outburst on MJD~53944.}
\label{tab:MJD53944}
\centering
\begin{tabular}{c|ccccccc}
\hline\hline
Free parameters	&	flare1	&	flare2	&	flare3  &  flare4  &  flare5  &  flare6	&  Notes\\
\hline	
$\delta_{\rm D}$   & 20 & 20 & 20 & 20 & 20 & 20 & Doppler factor		\\
$r$ ($\times 10^{-3}$pc)    & 3.3 & 1.5 & 2.2 & 2.6 & 1.5 & 5.0 & Distance to the jet base		\\
$R^{\prime}$	($\times 10^{14}$cm)   & 2.67 & 1.21 & 1.78 & 2.11 & 1.21 & 4.05 & Blob's radius\\
$B^{\prime}$ ($\times 10^{-1}$G) & 0.97 & 2.10 & 1.46 & 1.23 & 2.10 & 0.64 & Magnetic field\\
$L^{\prime}_{\rm inj}$ ($\times10^{42}$erg/s)    & 2.86 & 1.15 & 2.70 & 3.05 & 1.80 & 3.80 & Electron injection luminosity\\
$s$ & 2.4 & 2.4 & 2.4 & 2.4 & 2.4 & 2.4 & Spectral index of the EED\\
$\gamma^{\prime}_{\rm min}$ ($\times10^{4}$) & 7.5 & 7.5 & 7.5 & 7.5 & 7.5 & 7.5 & Minimum Lorentz factor\\
$\gamma^{\prime}_{\rm max}$ ($\times10^{6}$) & 2.0 & 2.0 & 2.0 & 2.0 & 2.0 & 2.0 & Maximum Lorentz factor\\
\hline
\end{tabular}
\end{minipage}
\end{table*}

\begin{figure}
\centering
\subfigure{
\includegraphics[width=0.9\columnwidth]{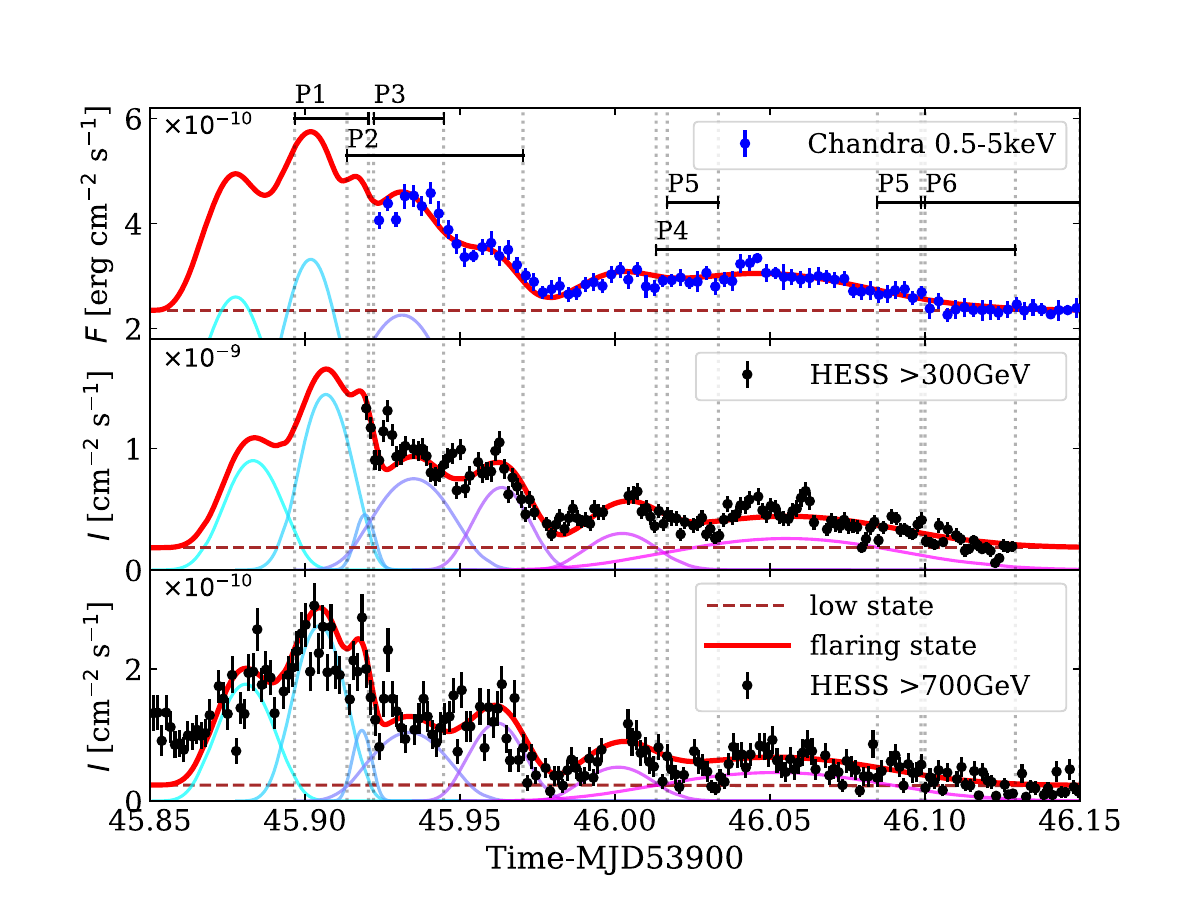}
}\hspace{-5mm}
\quad
\caption{Simulated multi-band LCs during the flare on MJD~53946. The upper panel shows the flux in X-ray band ranging from 0.5\,keV to 5\,keV, while the middle and lower panel display the $\gamma$-ray LCs with different energy thresholds, 300\,GeV and 700\,GeV respectively. The line styles have the same meaning as the left panel in Fig.~\ref{fig:MJD53944}. Besides, several time periods are displayed and the related information is listed in Table.~\ref{tab:period 53946}.
\label{fig:MJD53946 LC}}
\end{figure}

\begin{figure*}
\centering
\subfigure{
\includegraphics[width=0.9\columnwidth]{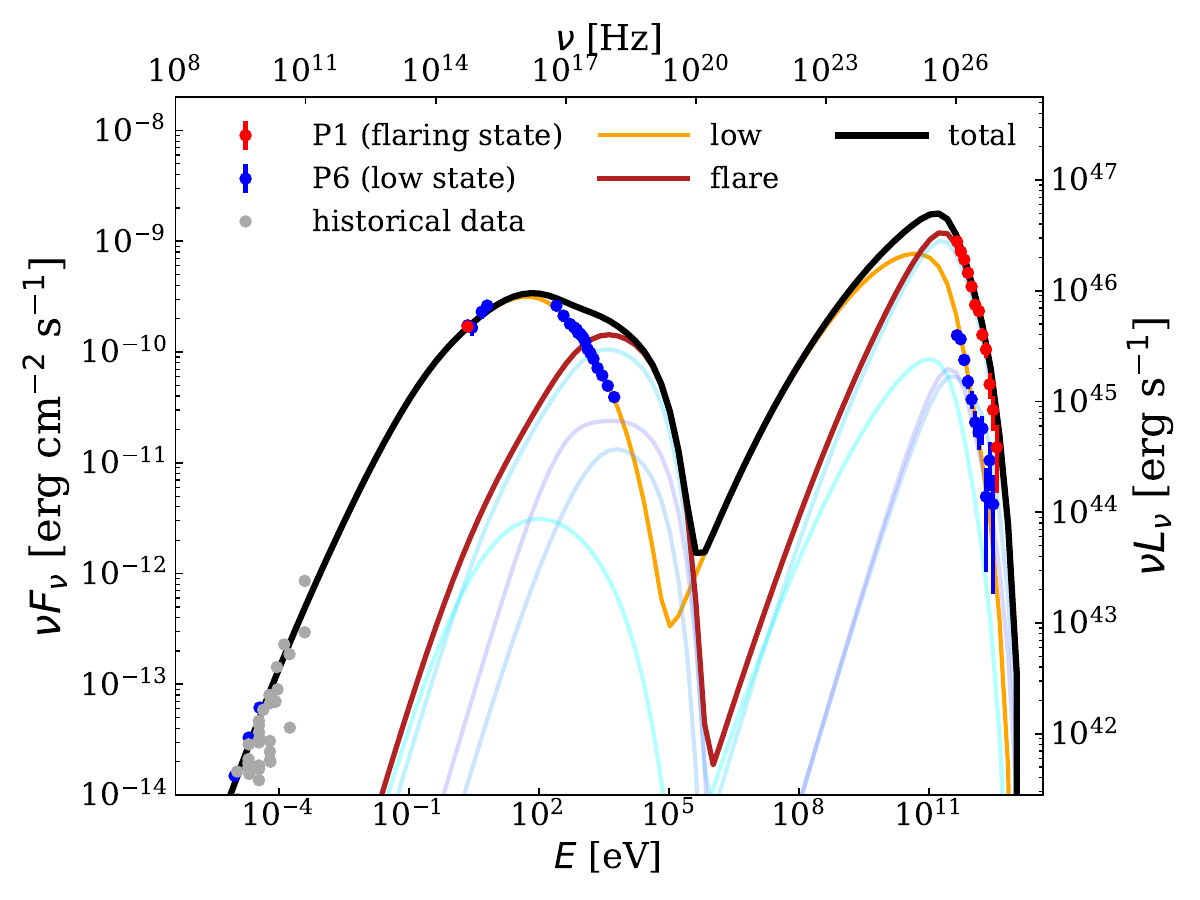}
}\hspace{-5mm}
\quad
\subfigure{
\includegraphics[width=0.9\columnwidth]{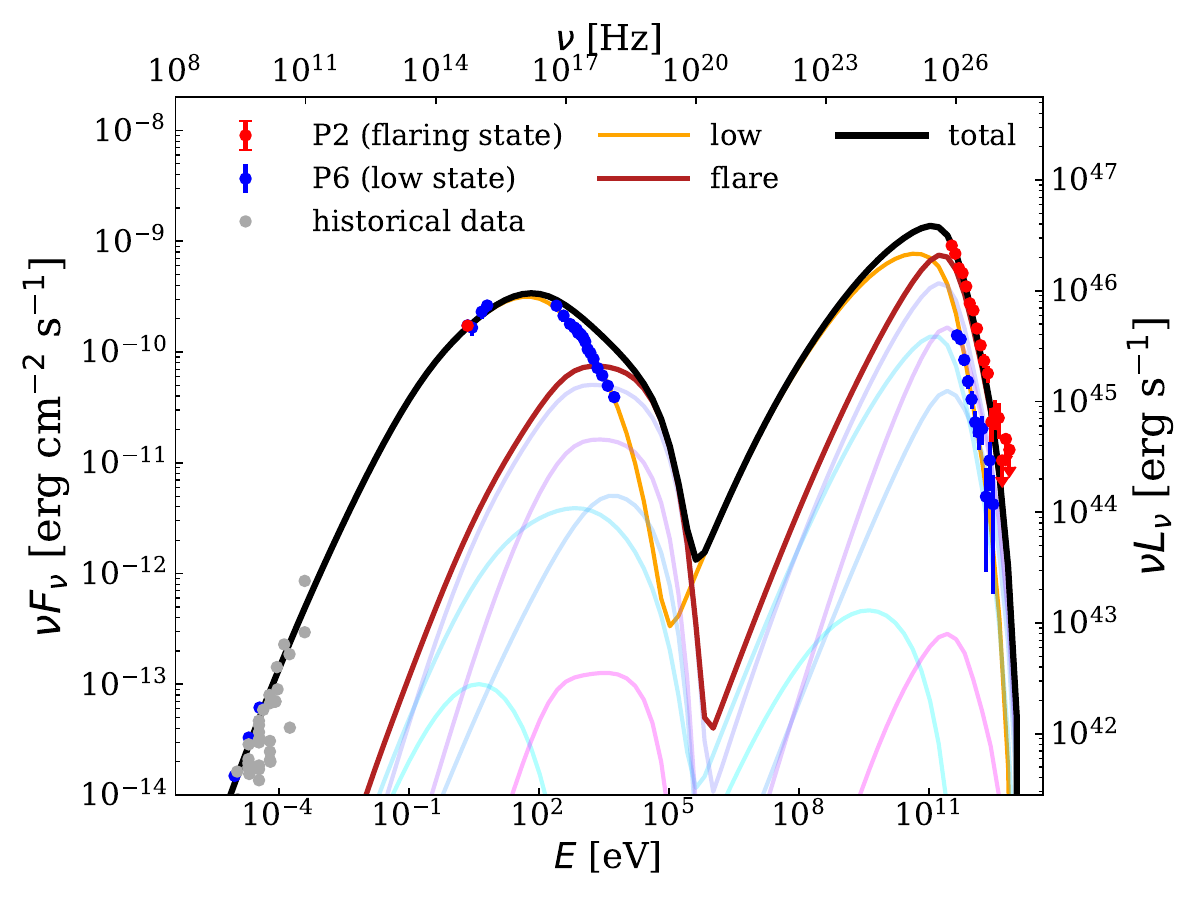}
}\hspace{-5mm}
\quad
\subfigure{
\includegraphics[width=0.9\columnwidth]{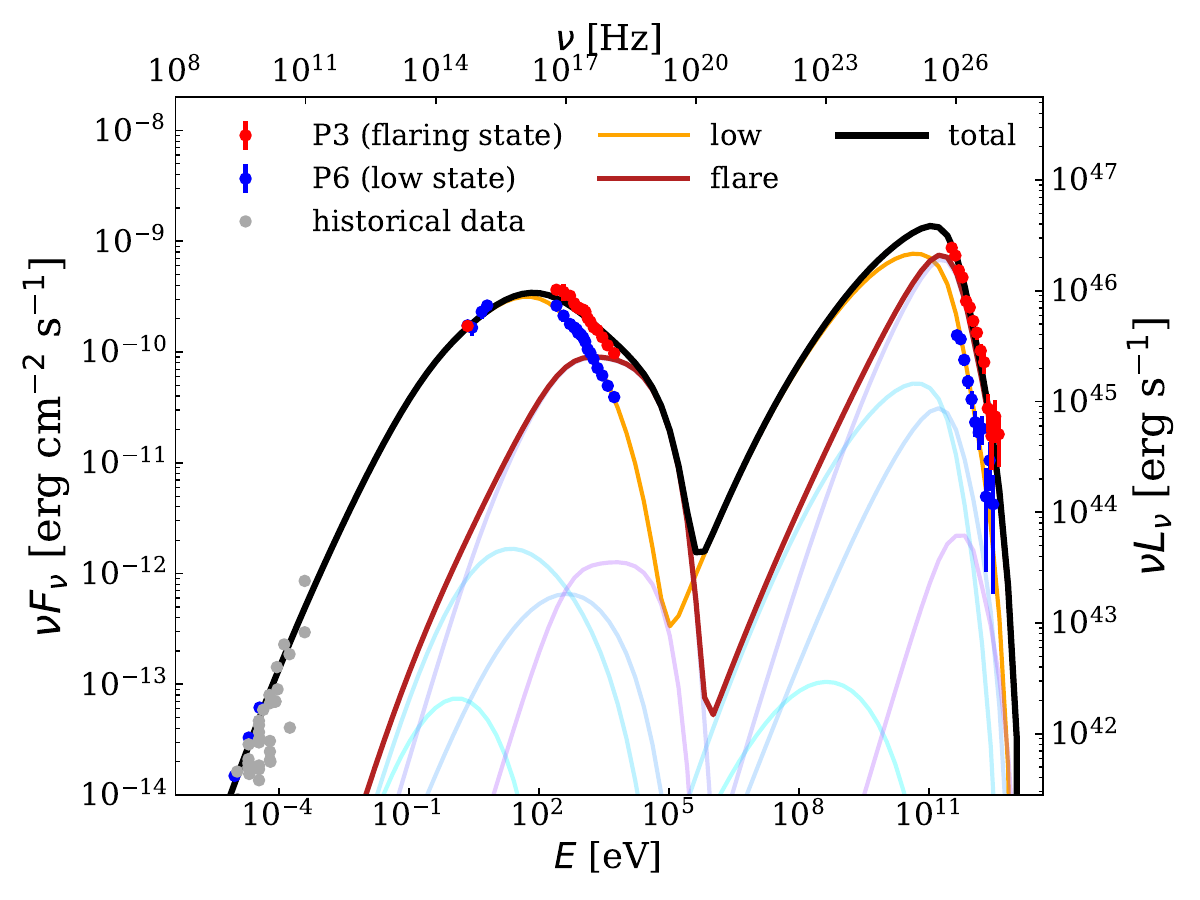}
}\hspace{-5mm}
\quad
\subfigure{
\includegraphics[width=0.9\columnwidth]{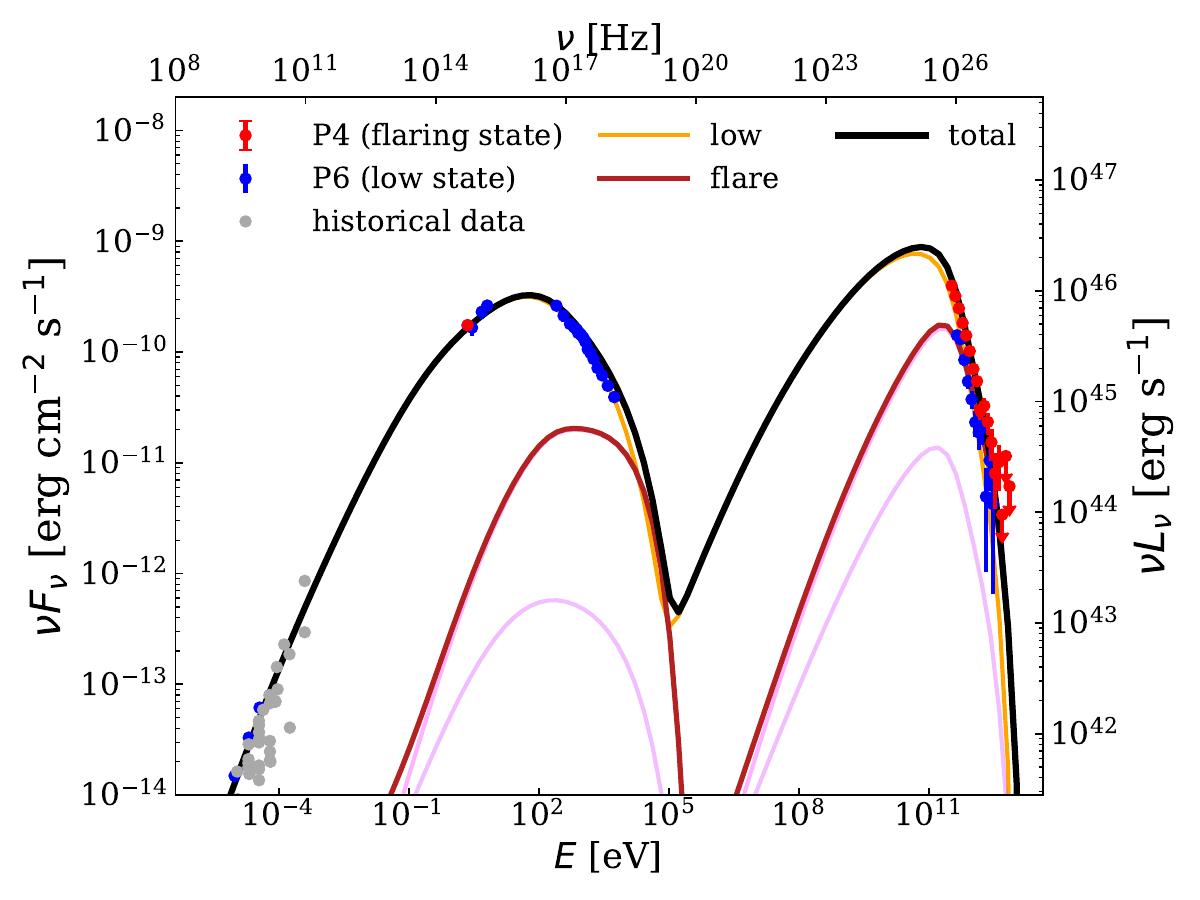}
}\hspace{-5mm}
\quad
\subfigure{
\includegraphics[width=0.9\columnwidth]{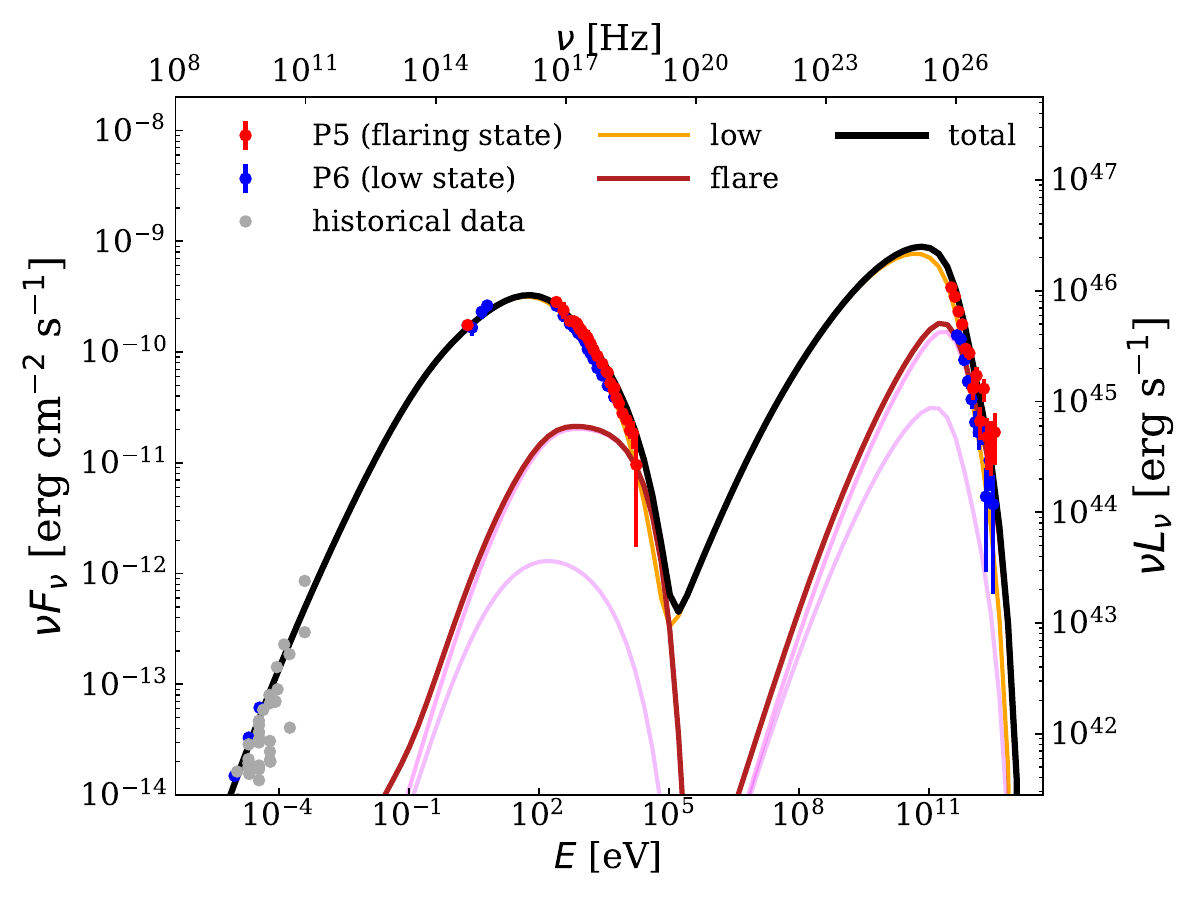}
}\hspace{-5mm}
\quad
\subfigure{
\includegraphics[width=0.9\columnwidth]{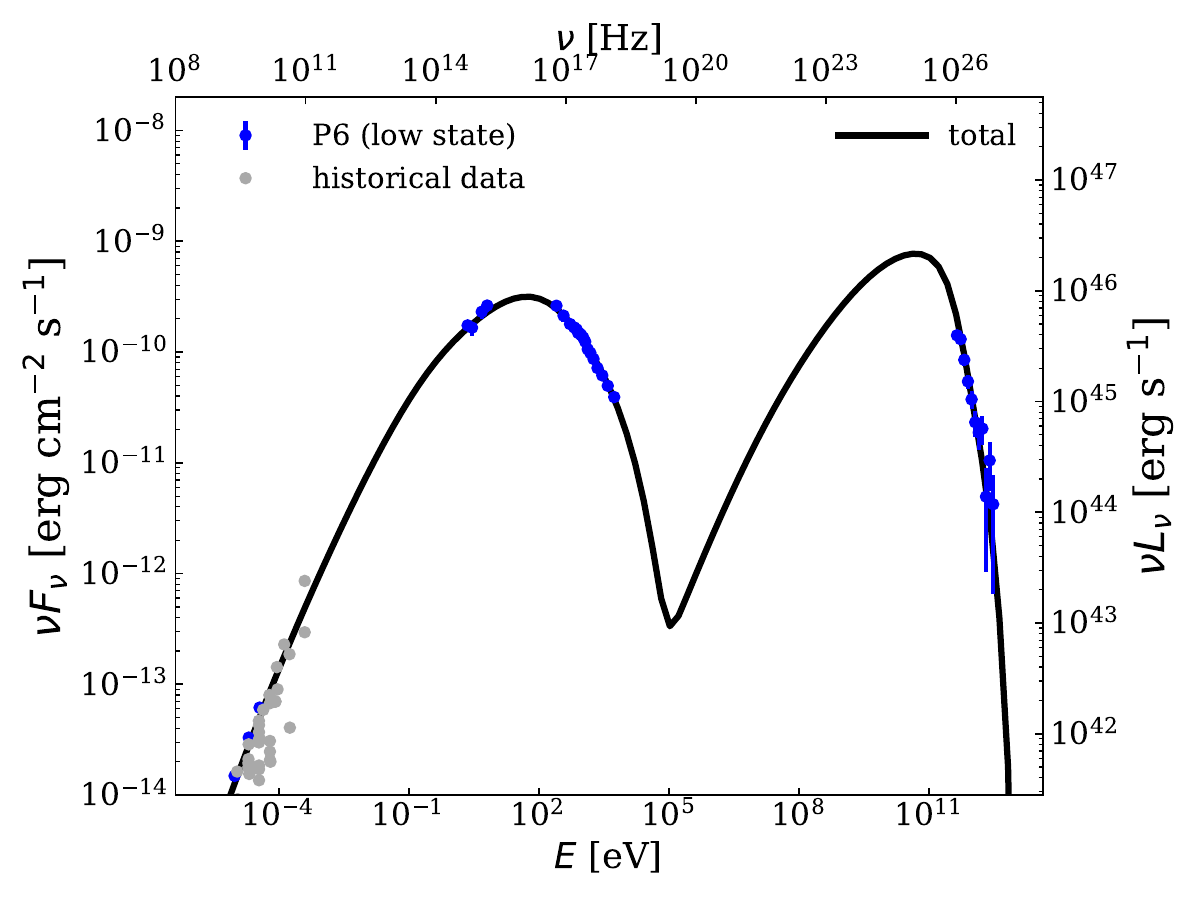}
}
\caption{Different average SEDs calculated in several time periods shown in the left panel of Fig.~\ref{fig:MJD53946 LC}. The line styles have the same meaning as the right panel of Fig.~\ref{fig:MJD53944}. 
\label{fig:MJD53946 SED}}
\end{figure*}

\begin{figure*}
\centering
\subfigure{
\includegraphics[width=0.68\columnwidth]{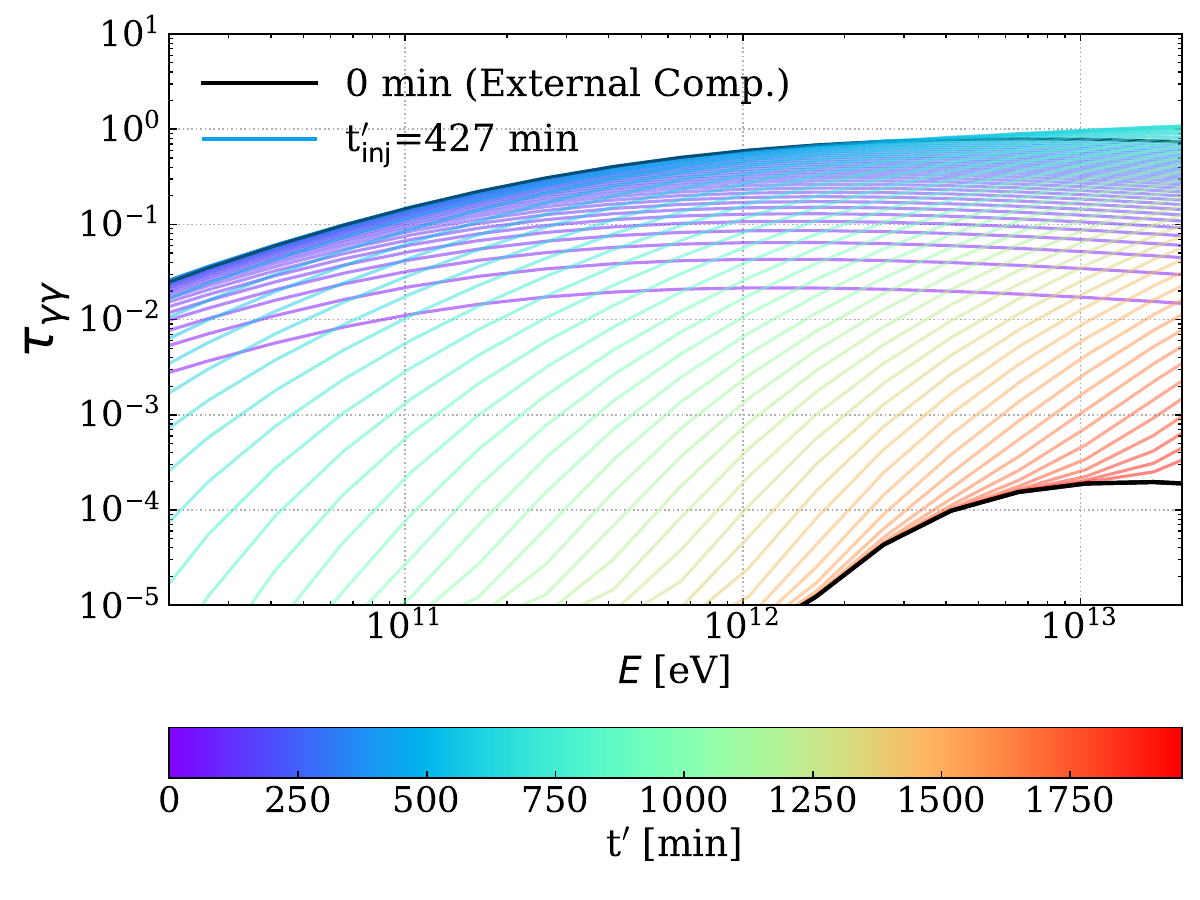}
}\hspace{-5mm}
\quad
\subfigure{
\includegraphics[width=0.68\columnwidth]{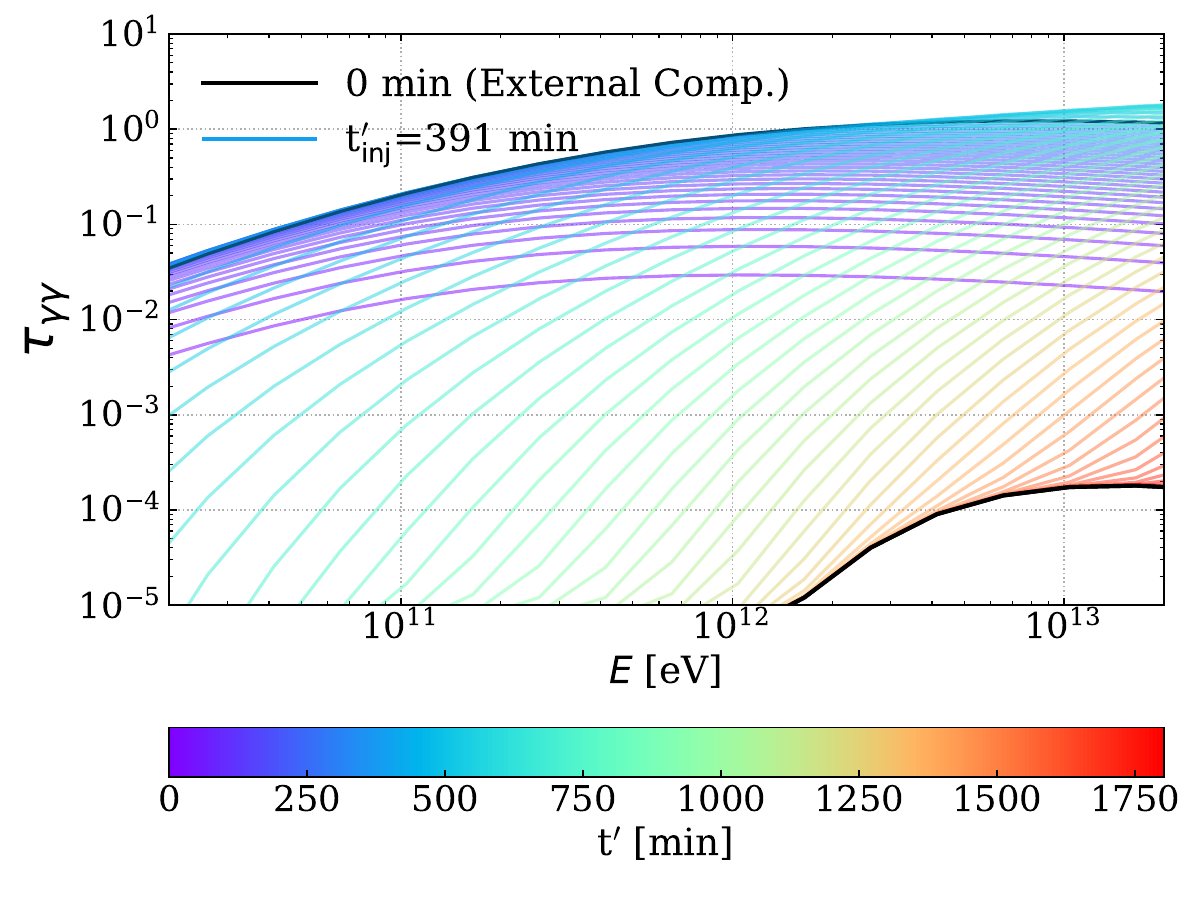}
}\hspace{-5mm}
\quad
\subfigure{
\includegraphics[width=0.68\columnwidth]{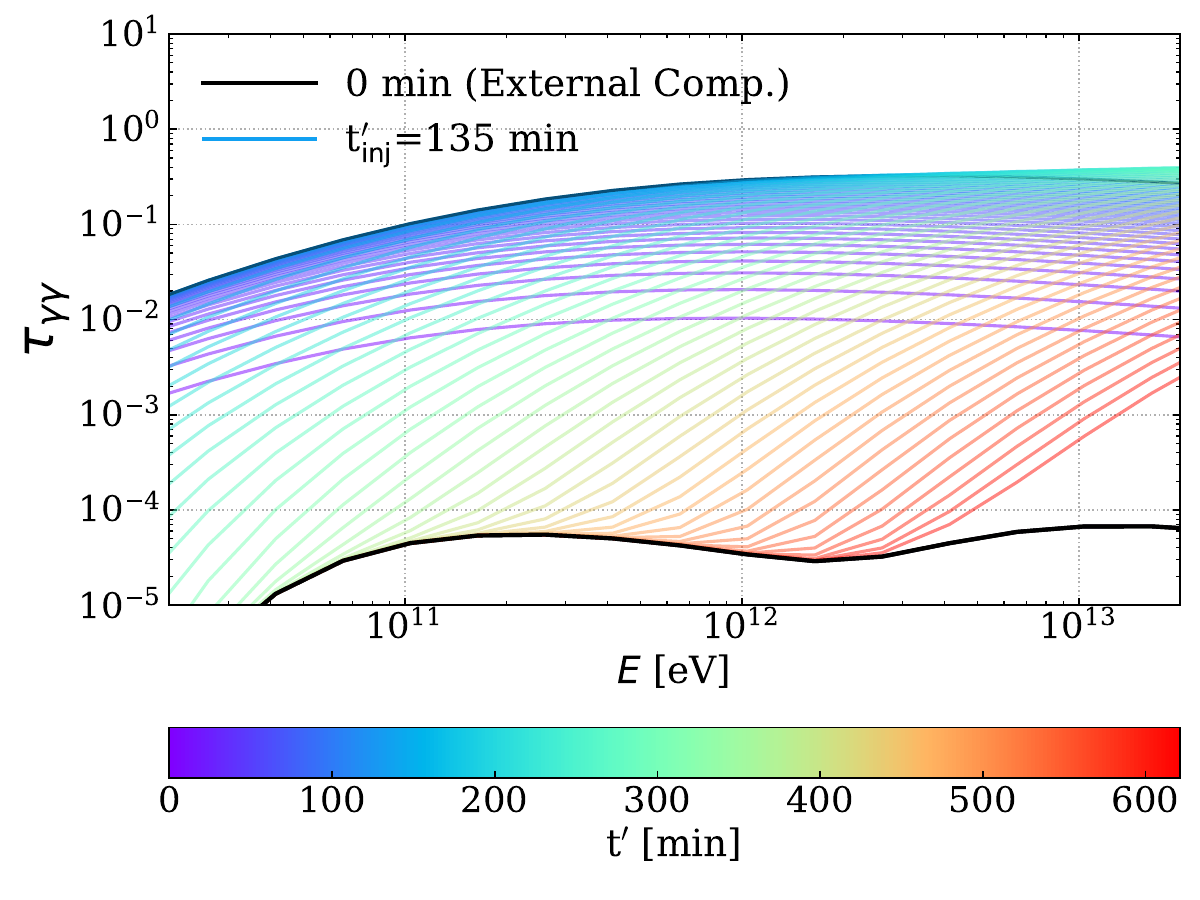}
}\hspace{-5mm}
\quad
\subfigure{
\includegraphics[width=0.68\columnwidth]{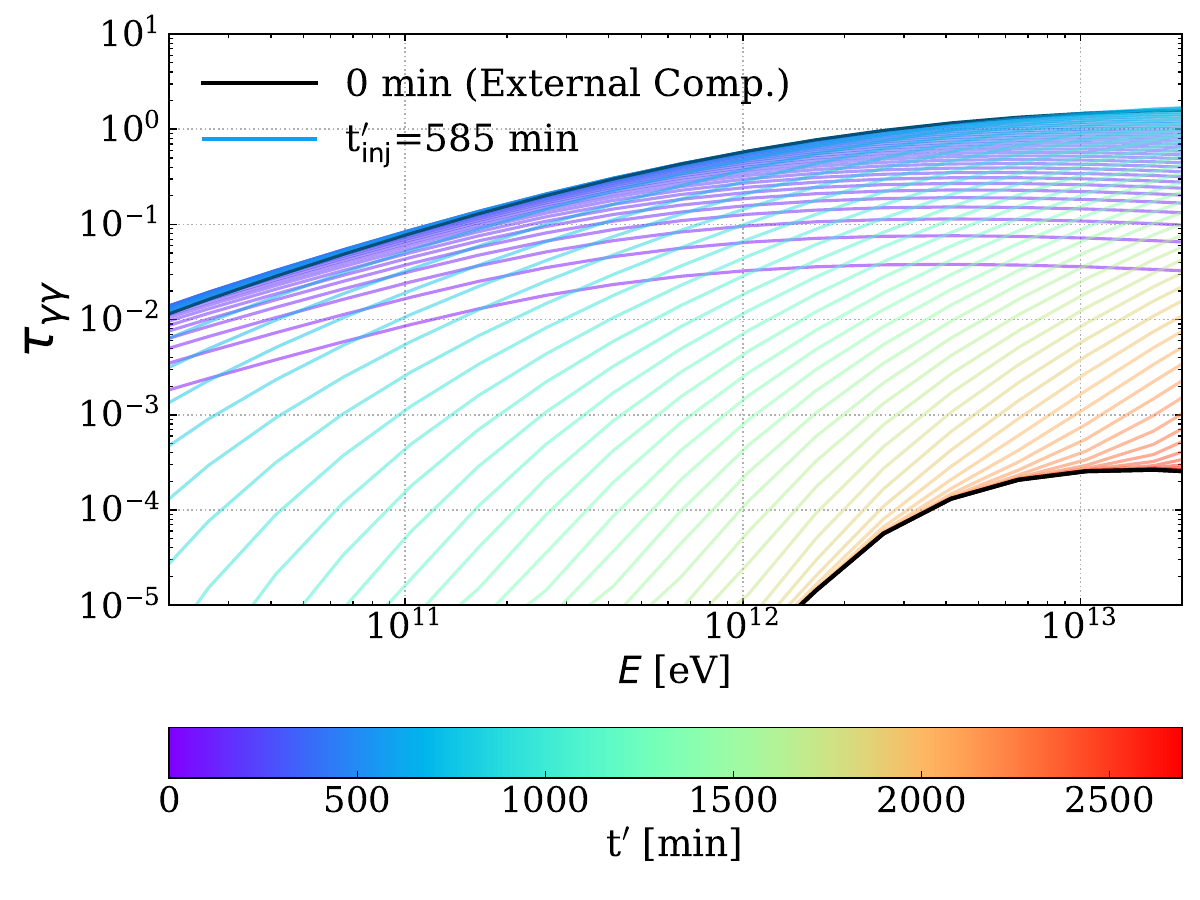}
}\hspace{-5mm}
\quad
\subfigure{
\includegraphics[width=0.68\columnwidth]{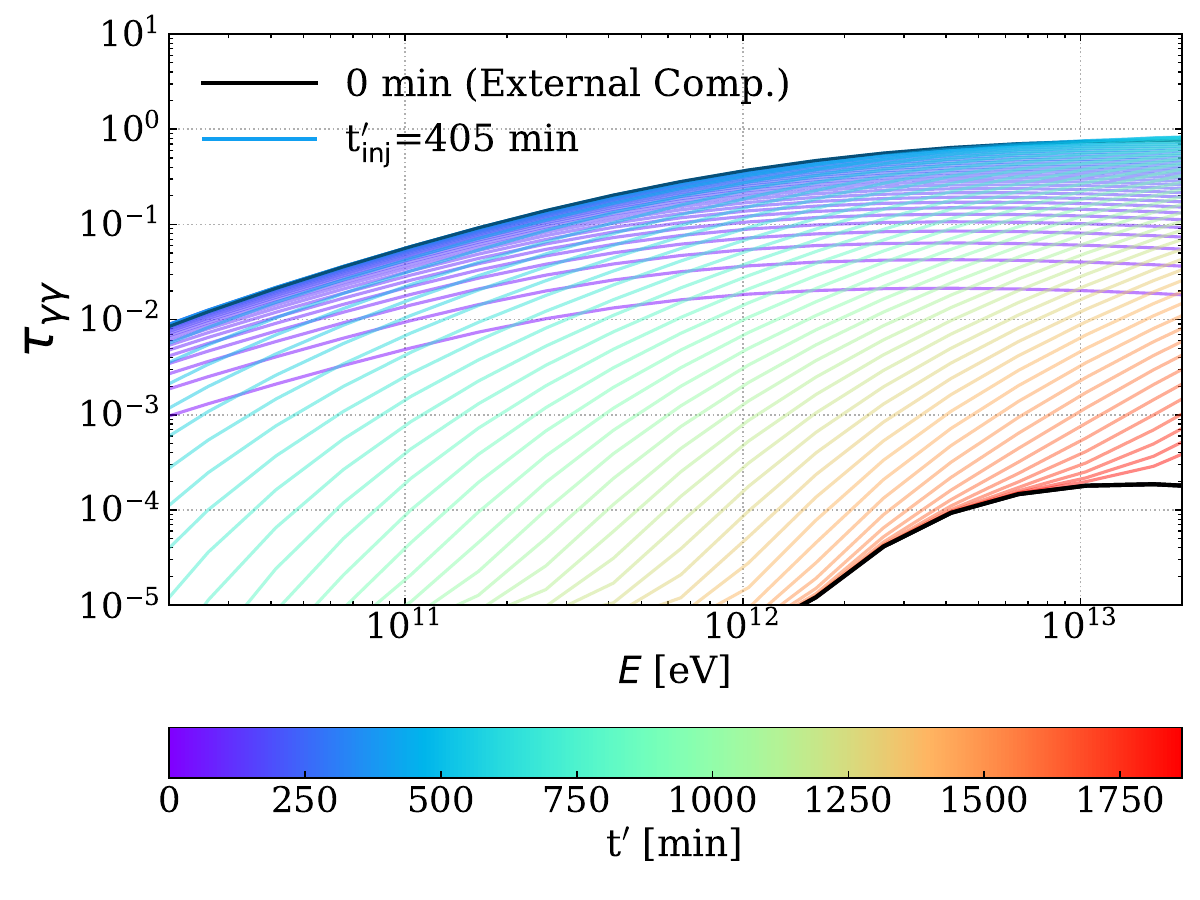}
}\hspace{-5mm}
\quad
\subfigure{
\includegraphics[width=0.68\columnwidth]{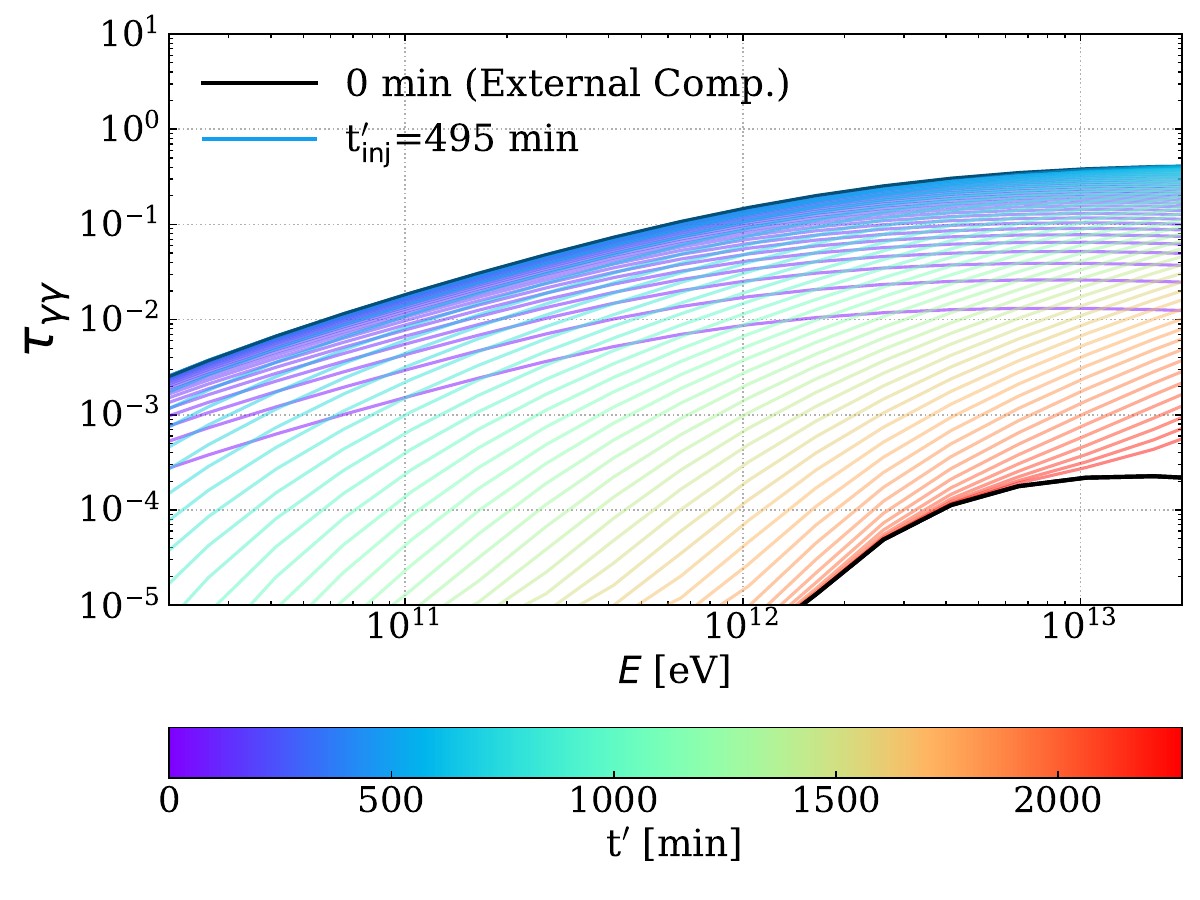}
}
\quad
\subfigure{
\includegraphics[width=0.68\columnwidth]{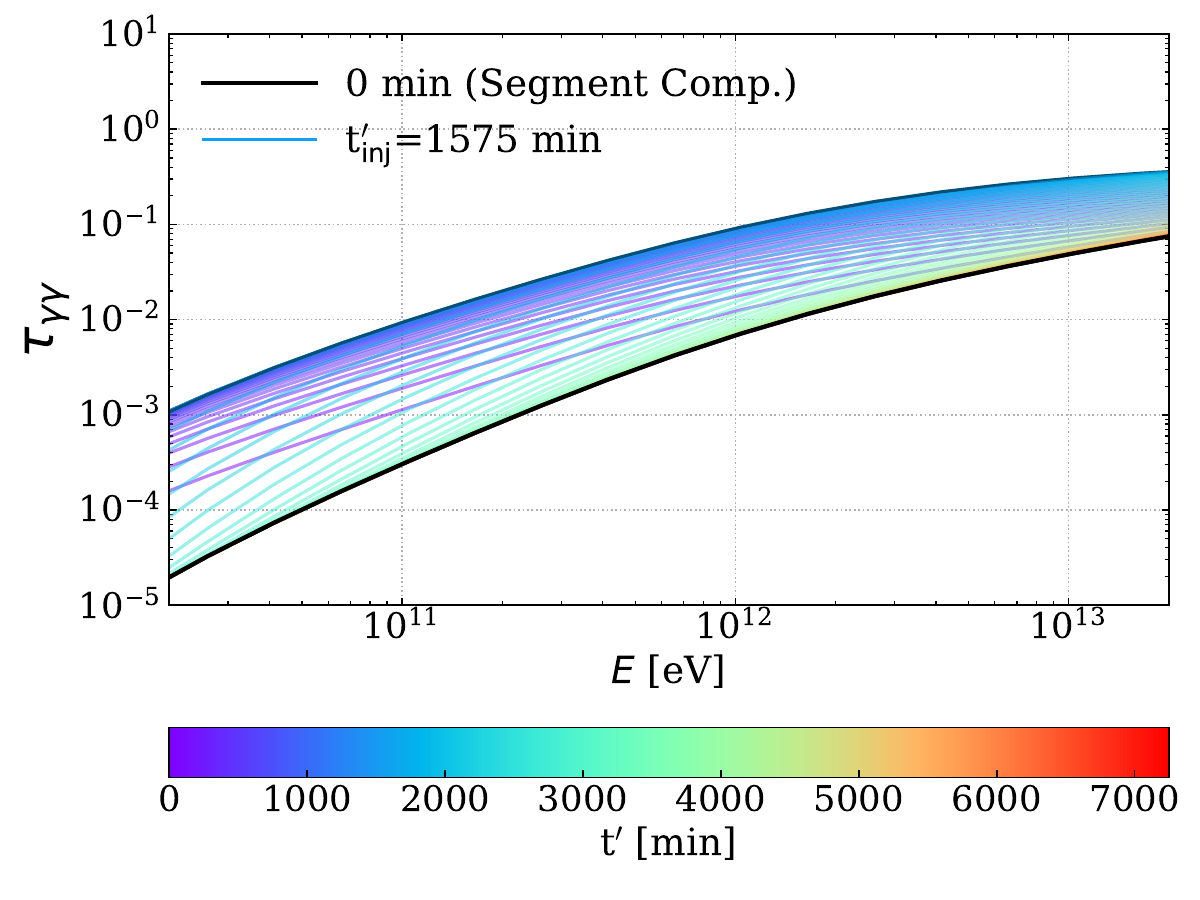}
}
\caption{Energy-dependent opacity of $\gamma\gamma$ absorption of each blob during the outburst on MJD~53946. The line styles have the same meaning as Fig.~\ref{fig:MJD53944 tau}.
\label{fig:MJD53946 tau}}
\end{figure*}

\begin{table*}
\begin{minipage}[t][]{\textwidth}
\caption{Summary of Parameters for the Outburst on MJD~53946.}
\label{tab:MJD53946}
\centering
\begin{tabular}{c|cccccccc}

\hline\hline
Free parameters	&	flare1	&	flare2	&	flare3  &  flare4  &  flare5  &  flare6	&  flare7 &  Notes\\
\hline	
$\delta_{\rm D}$   & 20 & 20 & 20 & 20 & 20 & 20 & 20 & Doppler factor		\\
$r$ ($\times 10^{-3}$pc)    & 9.5 & 8.7 & 3.0 & 13.0 & 9.0 & 11.0 & 35.0  & Distance to the jet base		\\
$R^{\prime}$	($\times 10^{14}$cm)   & 7.69 & 7.05 & 2.43 & 10.5 & 7.29 & 8.91 & 28.3 & Blob's radius\\
$B^{\prime}$ ($\times 10^{-1}$G) & 1.10 & 1.23 & 0.80 & 1.50 & 0.88 & 0.58 & 0.44 & Magnetic field\\
$L^{\prime}_{\rm inj}$ ($\times10^{42}$erg/s)    & 2.70 & 3.30 & 2.30 & 1.80 & 2.20 & 1.80 & 1.70 & Electron injection luminosity\\
$s$ & 2.8 & 2.8 & 2.8 & 2.8 & 2.8 & 2.8 & 2.8 & Spectral index of the EED\\
$\gamma^{\prime}_{\rm min}$ ($\times10^{5}$) & 1.8 & 1.8 & 2.5 & 0.9 & 1.2 & 1.0 & 1.0 & Minimum Lorentz factor\\
$\gamma^{\prime}_{\rm max}$ ($\times10^{6}$) & 1.5 & 1.5 & 1.5 & 1.5 & 1.5 & 1.5 & 1.5 & Maximum Lorentz factor\\
\hline
\end{tabular}
\end{minipage}
\end{table*}

\begin{table}
\caption{Information of time periods on MJD~53946 and correspondence between time period naming in this paper and naming in the \citet{2009A&A...502..749A}.}
\label{tab:period 53946}
\centering
\begin{tabular}{c|ccc}
\hline\hline
Period & starting time & end time & corresponding \\
\ & (MJD-53946) & (MJD-53946) & period \\
\hline	
P1 & 45.896643 & 45.920474 & T400-Peak \\
P2 & 45.913530 & 45.970312 & T300-High \\
P3 & 45.922130 & 45.944699 & T300-Xmax \\
P4 & 46.013252 & 46.129166 & T300-Low \\
P5 & 46.016846 & 46.033328 & \ \\
\  & 46.084624 & 46.098698 & T300-RXTE \\
P6 & 46.100000 & 46.150000 & T400-Xmin \\
\hline
\end{tabular}
\end{table}

\subsection{Modeling the outburst on MJD~53946}

Multiwavelength observations were carried out during the outburst on MJD~53946, which revealed the activity in different energy bands. This provides more restrictions on the parameters selection. However, the time resolution of the gamma-ray data is worse (i.e., the time binning is broader), which makes the identification of flares in the LC more ambiguous. We employ seven consecutive flaring blobs to fit the data, with parameters of each flares listed in Table.~\ref{tab:MJD53946}. The comparison between the modeled multiwavelength LCs and observations is shown in Fig.~\ref{fig:MJD53946 LC}. In the figure, we label several time periods during which the SEDs of the blazar were given by \citet{2009A&A...502..749A}. Related information of these periods is listed in Table.~\ref{tab:period 53946}. Modeled SEDs for these periods are compared with observations in Fig.~\ref{fig:MJD53946 SED}. The Doppler factor of all the blobs is still fixed to be 20. Compared with the first outburst, the simultaneous X-ray data in the second outburst leads to better constraints on parameters such as the magnetic field and the electron injection luminosity. 

The evolution of $\tau_{\gamma\gamma}$ for each flare during the second outburst is displayed in Fig.~\ref{fig:MJD53946 tau}. Similar to that in the first outburst as shown in Fig.~\ref{fig:MJD53944 tau}, the opacity of 1\,TeV photon in each flare is less than unity, indicating an unimportant attenuation of the TeV flux inside the jet. 
It may be worth noting that, the flare 7 has a relatively long variability timescale, and hence the corresponding blob is located at a relatively large distance. The blob is embedded inside the region where numerous non-flaring blobs are generated. Hence the opacity caused by synchrotron radiation from nearby non-flaring blobs plays a more important role than the external thermal radiation, although the absorption is still quite weak. This is reflected in the last panel of Fig.~\ref{fig:MJD53946 tau}. 

\begin{figure}
\centering
\subfigure{
\includegraphics[width=0.9\columnwidth]{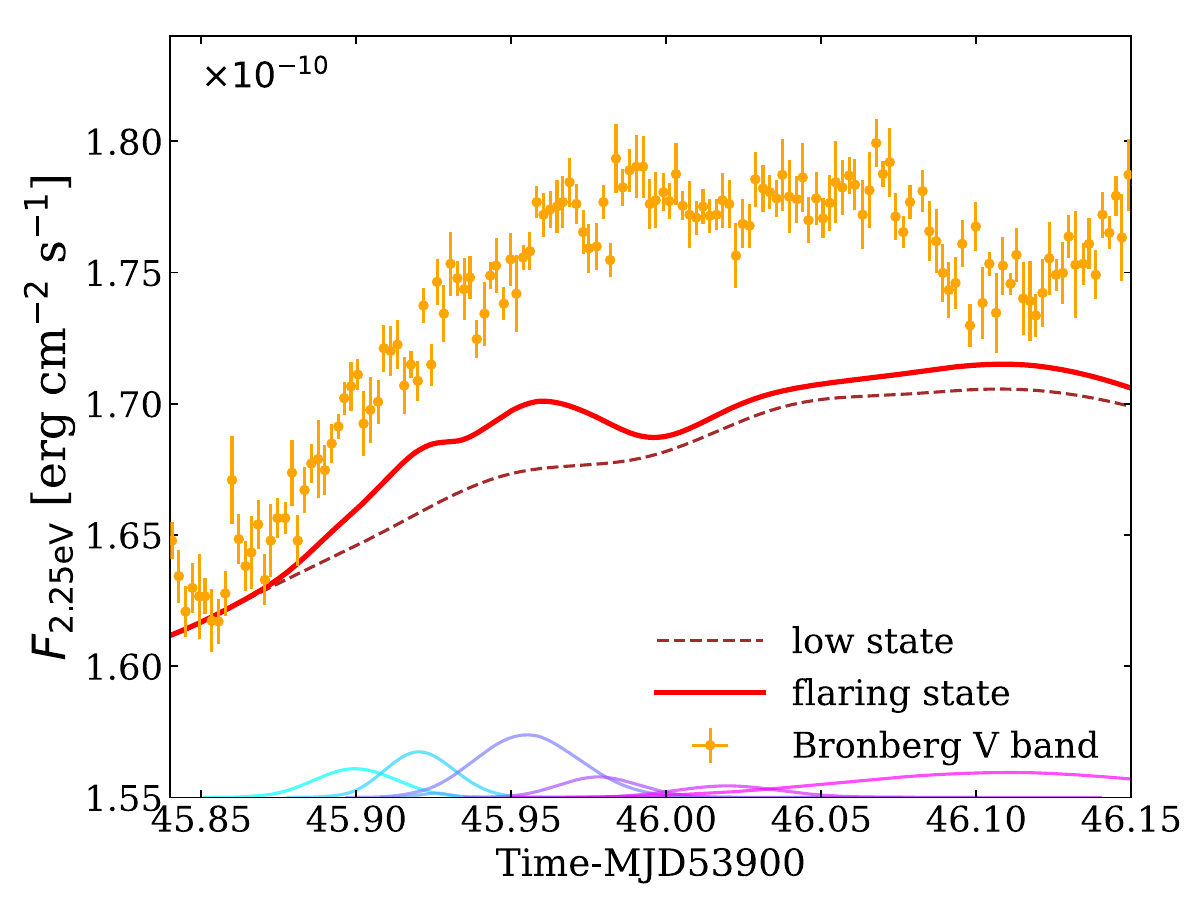}
}\hspace{-5mm}
\quad
\caption{Simulated optical LC during the flare on MJD~53946. The brown dashed line is the time-dependent optical flux of LS2 and the red solid line represents the total emission combining the emission of the low state and flaring component. The gradient color lines represent the optical flux of each blob. For visibility, a constant value is added to the flux of each blob, i.e., $\nu F_{\nu, \rm 2.25eV}+1.55\times10^{-10}$\,erg\,cm$^{-2}$\,s$^{-1}$.
\label{fig:MJD53946 LC2}}
\end{figure}

\section{Discussion}\label{sec:discussion}

\subsection{Identification of Flares}

\citet{2007ApJ...664L..71A} identified five flares in the $\gamma$-ray observations of the first outburst. We followed this result and assumed five blobs generated at differernt times. However, when modeling the LC with this setup, there would be a gap around 53~min (t-MJD~53944) in the predicted LC which does not fit the observation well. Therefore, we added another flaring blob (i.e., the flare 2 for the first outburst) to fill the gap. During the second outburst, five flares were suggested by \citet{2009A&A...502..749A} based on the $\gamma$-ray flux (corresponding to flare 1, 2, 3, 5, 7 in our model for the second outburst). However, with only these five flares our model cannot reproduce some structures in X-ray and/or $\gamma$-ray band, such as the bump around $\sim$MJD~53945.935 and another minor one peaking at $\sim$MJD~53946.02. We therefore added two other flaring blobs in the model, namely, flare 4 and 6 of the second outburst, to account for the LC. It should be noted that the number of flaring blobs need in a theoretical model depends on both the LCs of individual blobs predicted by the model and the quality of the observations. For example, flare 2 of the first outburst may not be needed if the model predicted a shallower decay phase of flare 1. On the other hand, poor statistics of the data would make the identification of flares ambiguous. In the second outburst, the size of the time bins of the $>700\,$GeV LC is twice that of the first outburst, which thus reveals less short-term structure in the LC. Besides, the larger error bars and larger variation in its LC also make it difficult to distinguish between statistical fluctuations and true flares. Therefore, our modelling in this work should only be considered a schematic description of the temporal behaviour of the outbursts, from which no detailed conclusions about exact parameter values should be drawn.

\subsection{Optical emission}

Optical observations of PKS~2155-304 were carried out by the Bronberg Observatory during the second outburst. The optical flux underwent a rising phase almost at the same time as the $\gamma$-ray flux, and reached a plateau with a flux enhancement of only 15~\% with respect to the low state. Small variations can be seen in the optical LC, but they are not correlated with the X-ray and $\gamma$-ray variability. The flaring blobs do not contribute significantly to the optical flux due to a relatively large $\gamma_{\rm min}'(\sim 10^5)$ employed. We therefore test whether the small fluctuations of the low-state emission (from non-flaring blobs) can account for the optical variability. Therefore, instead of using the long-term average flux (Eq.~\ref{eq:average}), we adopt Eq.~(\ref{eq:lowstate}) for the low-state emission during the second outburst. The red solid curve in Fig.~\ref{fig:MJD53946 LC2} shows the sum of the flux of the flaring blobs during the second outburst and the low-state emission. The total flux is still not sufficient to reproduce the observed flux in the optical band. Nevertheless, the deficiency of the flux is only about 5 -- 6~\%, so it does not affect the overall fitting of the SED as shown in Fig.~\ref{fig:MJD53946 SED}.   
As optical flux measurements may be affected by systematic errors of the order of $\sim 10$~\%, we may still consider our fit to be a reasonable representation of the data. 
An even better fit could possibly be achieved by accounting for fluctuations of physical properties in those non-flaring blobs at a relatively large distance. A slightly higher electron injection luminosity would fill the flux deficiency. Alternatively, the additional optical flux could be generated by an extra electron component with different origin from the X-ray and gamma-ray emitting electrons, such as a long-time continuous injection \citep[e.g.][]{2008MNRAS.390L..73B}.

\subsection{Model comparison}

Efforts have been made to model the emission of the two exceptional outbursts of PKS~2155-304 with a moderate Doppler factor under the multi-zone framework since they were reported. \citet{2008MNRAS.390..371K} proposed that the emission during the outburst arises from a relatively large blob, which dominates the X-ray emission, and some compact blobs responsible for individual $\gamma$-ray flares \citep[see also][]{2012A&A...539A.149H}. The model reproduced the minute-scale TeV variability with a moderate Doppler factor of $20-30$. The reason is similar to what was discussed in the previous section: the soft X-ray radiation of the blazar, which is the main absorber of TeV photons, is not entirely from the same radiation zone as the TeV flares. The main difference lies in the fact that the extended jet considered in our model is replaced with a large blob. Therefore, the radio emission is unaccounted for in their model. Note that in our model the radio emission can provide important constraints on the properties of the jet, which can consequently affect the broadband radiation of the blazar. 
As an alternative, \citet{2001A&A...367..809K, 2003A&A...410..101K} established a blob-in-jet model to interpret the multiwavelength variability of blazars. In that model, an inner jet is assumed to account for the blazar's radiation from the radio to the optical band, or even up to the ultraviolet band, while the X-ray and $\gamma$-ray emission is ascribed to compact blobs. Sometimes, extended radio structures, such as ``radio clouds'' need to be included to explain the low-frequency radio emission. \citet{2012A&A...539A.149H} adopted that model to explain the X-ray and $\gamma$-ray observations of the second outburst. The significant enhancements of the X-ray and $\gamma$-ray fluxes during the outburst are attributed to the sudden injection or acceleration of electrons inside some compact blobs. Since the X-ray emission and the gamma-ray emissions are from the same region, this model suffers the same problem of strong attenuation of TeV photons as the classical one-zone model, leading to the "Doppler factor crisis". Indeed, a Doppler factor as high as 50 is employed in the modelling \citep{2012A&A...539A.149H}.

\begin{figure}
\subfigure{
\centering
\includegraphics[width=0.9\columnwidth]{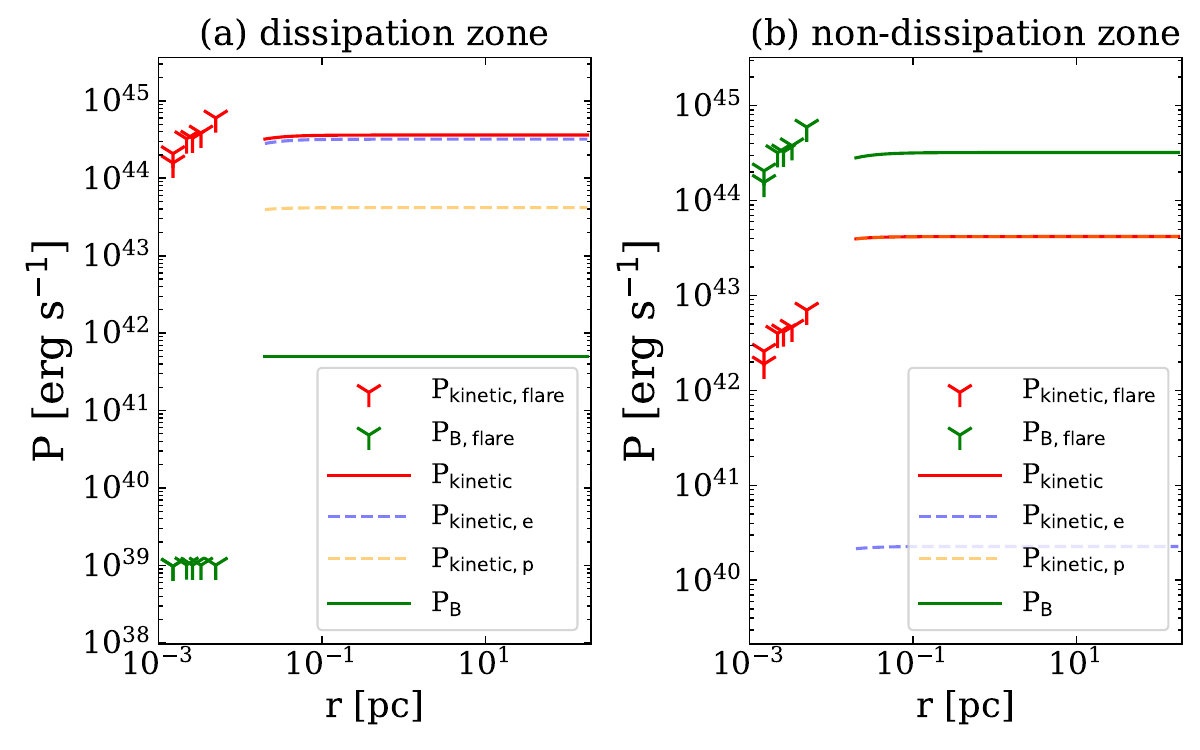}
}\hspace{-5mm}
\quad
\subfigure{
\centering
\includegraphics[width=0.9\columnwidth]{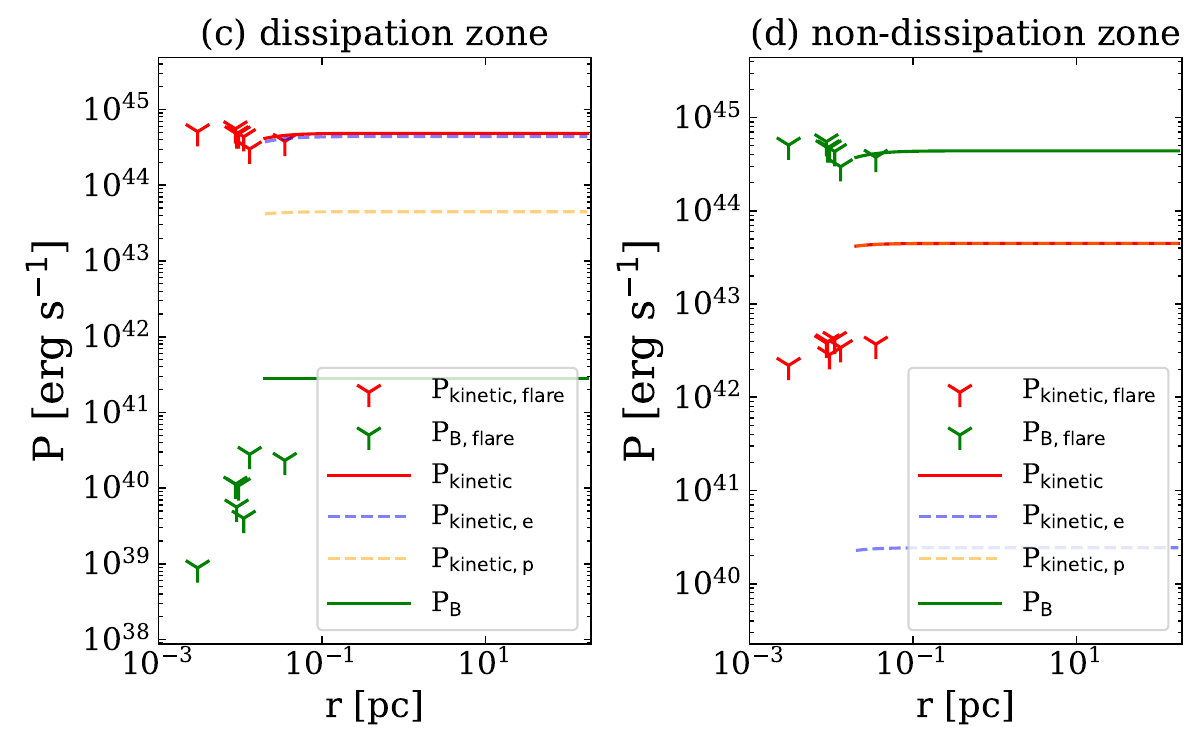}
}\hspace{-5mm}
\quad
\caption{Panel(a,b) display the kinetic and Poynting flux of the dissipation and non-dissipation zone in LS1 and the first outburst while panel(c,d) show the result in LS2 and the second outburst.
\label{fig:Power}}
\end{figure}

\subsection{Jet power}

The average differential electron number density distribution within a segment at a distance of $r$ from the SMBH is given by
\begin{equation}
    \frac{dn^{\prime}_{e}}{d\gamma^{\prime}} = N_{\rm blob}(r) Q^{\prime}(r,\gamma^{\prime})t^{\prime}_{\rm equ}(r,\gamma^{\prime}),
\end{equation}
where $t^{\prime}_{\rm equ}(r,\gamma^{\prime}) = \left[t^{\prime -1}_{\rm cool}(r,\gamma^{\prime})+t^{\prime -1}_{\rm ad}(r)+t^{\prime -1}_{\rm esc}(r)\right]^{-1}$. The electron number density can be calculated as $n_e=\int{\frac{dn^{\prime}_{e}}{d\gamma^{\prime}}d\gamma^{\prime}}$. Assuming that the dissipation is triggered by a magnetic reconnection event, the energy of accelerated electrons is converted from the magnetic field energy. If the jet is composed of equal amounts of protons and electrons, the kinetic power can be estimated by
\begin{equation}
    \begin{split}
        &P_{\rm k,0}(r) = \pi R^{\prime 2}_{\rm jet}(r) \Gamma^2 c n_e (m_e+m_p) c^2, \\
        &P_{\rm k}(r) = \pi R^{\prime 2}_{\rm jet}(r) \Gamma^2 c \left(m_e c^2 \int{\gamma^{\prime}\frac{dn^{\prime}_{e}}{d\gamma^{\prime}}d\gamma^{\prime}} + n_e m_p c^2\right), 
    \end{split}
\end{equation}
where $P_{\rm k,0}(r)$ and $P_{\rm k}(r)$ are the kinetic powers in the non-dissipation and dissipation zone, respectively and we only consider electron acceleration. The Poynting flux can be given by
\begin{equation}
    \begin{split}
        &P_{B,0}(r) = \pi R^{\prime 2}_{\rm jet}(r) \Gamma^2 c \left( u^{\prime}_{B} +  m_e c^2 \int{(\gamma^{\prime}-1)\frac{dn^{\prime}_{e}}{d\gamma^{\prime}}d\gamma^{\prime}}\right), \\
        &P_{B}(r) = \pi R^{\prime 2}_{\rm jet}(r) \Gamma^2 c u^{\prime}_{B}.
    \end{split}
\end{equation}
We then substitute our model parameters into above equations. The obtained Poynting and kinetic fluxes of the dissipation and non-dissipation zones in the two low states and two outburst are shown in Fig.~\ref{fig:Power}. The Poynting-to-kinetic flux ratios are approximately $10^{-3}$ and $10$ in the dissipation and non-dissipation zone beyond $r_0=0.02$~pc in the jet. Since our model considers discrete dissipation zones embedded in the jet, we need to average over the dissipating regions and the non-dissipating regions of the jet to estimate the overall Poynting-to-kinetic flux ratios. Given the filling factor $\lambda$ of $60\%$, the average ratio is estimated to be $\sigma_{\rm ave}=(\lambda^{2/3}P_B+(1-\lambda^{2/3})P_{B,0})/(\lambda^{2/3}P_{\rm k}+(1-\lambda^{2/3})P_{\rm k,0})\sim0.35$. 
Therefore, in our model, the jet is magnetically dominated with a magnetization $\sigma \gtrsim 10$ at $r<r_0=0.02\,$pc, where dissipation zones are not stably formed, and becomes matter-dominated in the outer region where dissipation becomes important and the magnetization decreases ($\sigma \sim 0.35$). This is consistent with previous simulations where an initial magnetization parameter of $10-100$ is obtained at the jet base \citep[e.g.][]{2007MNRAS.380...51K} and decreases along the jet \citep[e.g.][]{2019MNRAS.490.2200C} in the theoretical framework of magnetically driven jets.

The total jet power is $P_{\rm jet}=P_{\rm k} + P_{B}$. In the two low states, the jet power is comparable, with $P_{\rm jet,LS1}=3.6\times10^{44} \rm erg~s^{-1}$ and $P_{\rm jet,LS2}=4.8\times10^{44} \rm erg~s^{-1}$. During the outbursts, the jet power is elevated to $P_{\rm jet,flare1}=4.8\times10^{44} \rm erg~s^{-1}$ and $P_{\rm jet,flare2}=6.7\times10^{44} \rm erg ~s^{-1}$, with an average enhancement by $30\%$. The ten-fold flux boost during the outbursts is mainly due to a higher SSC radiation efficiency in the compact dissipation zones formed at $r<r_0$. The obtained jet power is consistent with typical ranges of the jet power of blazars \citep{2008MNRAS.385..283C}. In previous models for the minute-scale TeV flare of PKS~2155-304, such as the one-zone SSC/EC model and the spine-layer jet model \citep{2008MNRAS.386L..28G}, the required jet power varies from $10^{43} \rm erg ~s^{-1}$ to $10^{45} \rm erg ~s^{-1}$, which is in line with our calculation.

On the other hand, if the dissipation mechanism is assumed to be internal collisions between different parts of the jet, the energy of the nonthermal electrons during the dissipation should originate from the kinetic energy of the jet before collision. However, given that the kinetic power (including nonthermal energy) of electrons is about 10 times the kinetic power of protons in the dissipation zone (see Fig.~\ref{fig:Power}), this would imply the original jet Lorentz factor to be also 10 times larger, i.e, $\sim 100$, which is unreasonably high. Therefore, our model prefers magnetic reconnection as the energization mechanism of nonthermal particles in the jet.

\section{Conclusions}\label{sec:conclusion}

The two gigantic outbursts of PKS~2155-304 discovered on 28 -- 30 July 2006 present extremely short timescales of sub-flares and remarkable flux enhancements in the TeV $\gamma$-ray band. This implies the existence of a compact high-energy emission region in the blazar jet. In the classical one-zone framework, a large Doppler factor is required to avoid absorption of TeV photons inside the compact emission region, leading to the so called ``Doppler factor crisis''. In this study, we modeled the SEDs and LCs of the two outbursts of PKS~2155-304 within a multi-blob framework, the stochastic dissipation model.  

In this model, the emission of the blazar during the outburst is composed of a quasi-stable low-state component which is the superposition of numerous non-flaring blobs along the jet, and a flaring component produced by a series of flaring blobs which are formed in strong dissipation events in the jet. The model can reproduce the minute-scale TeV variability with a moderate and typical value of $\delta_{\rm D}=20$ for the blazar jet, and explain the simultaneous X-ray observations during the second outburst. The opacity due to $\gamma\gamma$ absorption is significantly reduced compared to that in the one-zone model. This is mainly because the ultraviolet -- X-ray radiation, which serves as target photons for the attenuation process, is shared among many blobs along the jet, and consequently the target photon density is significantly reduced in a single flaring blob responsible for the TeV flare. Our model provides a solution to the ``Doppler factor crisis''. 

Our SED and LC fits require a significantly increased electron injection luminosity, increased magnetic field in a single blob compared to the parameters for the low state. The required jet's power $P_{\rm jet}$ is several times $10^{44}\rm~erg ~s^{-1}$ for both the low state and the flaring state, with the latter being about 30\% higher than the former. This is because the number of blobs (or dissipation zones) in the low state is more numerous. The TeV flux enhancement during the two outbursts may then be understood as the formation of compact and intense dissipation zones at small distance, leading to a significant boost of the IC radiation efficiency. The average magnetization parameter of the jet, i.e., Poynting-to-kinetic luminosity ratio, is approximately 0.3 at $r>r_0(=0.02\,$pc), where dissipation events can frequently and stably occur. Assuming the  dissipation is due to magnetic reconnection, we can derive the magnetization parameter of non-dissipating region of the jet is about 10, which is consistent with a magnetically driven jet.


\section*{Acknowledgements}
This work is supported by NSFC under the grant No.2031105.

\section*{Data Availability}

No new data were generated or analysed in support of this research.




\begin{thebibliography}{}
\bibitem[\protect\citeauthoryear{Acciari et al.}{2011}]{2011ApJ...738...25A} Acciari V.~A., Aliu E., Arlen T., Aune T., Beilicke M., Benbow W., Boltuch D., et al., 2011, ApJ, 738, 25. doi:10.1088/0004-637X/738/1/25
\bibitem[\protect\citeauthoryear{Aharonian}{2000}]{2000NewA....5..377A} Aharonian F.~A., 2000, NewA, 5, 377. doi:10.1016/S1384-1076(00)00039-7
\bibitem[\protect\citeauthoryear{Aharonian et al.}{2005a}]{2005A&A...430..865A} Aharonian F., Akhperjanian A.~G., Aye K.-M., Bazer-Bachi A.~R., Beilicke M., Benbow W., Berge D., et al., 2005, A\&A, 430, 865. doi:10.1051/0004-6361:20041853
\bibitem[\protect\citeauthoryear{Aharonian et al.}{2005b}]{2005A&A...442..895A} Aharonian F., Akhperjanian A.~G., Bazer-Bachi A.~R., Beilicke M., Benbow W., Berge D., Bernl{\"o}hr K., et al., 2005, A\&A, 442, 895. doi:10.1051/0004-6361:20053353
\bibitem[\protect\citeauthoryear{Aharonian et al.}{2007}]{2007ApJ...664L..71A} Aharonian F., Akhperjanian A.~G., Bazer-Bachi A.~R., Behera B., Beilicke M., Benbow W., Berge D., et al., 2007, ApJL, 664, L71. doi:10.1086/520635
\bibitem[\protect\citeauthoryear{Aharonian et al.}{2009}]{2009A&A...502..749A} Aharonian F., Akhperjanian A.~G., Anton G., Barres de Almeida U., Bazer-Bachi A.~R., Becherini Y., Behera B., et al., 2009, A\&A, 502, 749. doi:10.1051/0004-6361/200912128
\bibitem[\protect\citeauthoryear{Albert et al.}{2007}]{2007ApJ...669..862A} Albert J., Aliu E., Anderhub H., Antoranz P., Armada A., Baixeras C., Barrio J.~A., et al., 2007, ApJ, 669, 862. doi:10.1086/521382
\bibitem[\protect\citeauthoryear{Begelman, Fabian, \& Rees}{2008}]{2008MNRAS.384L..19B} Begelman M.~C., Fabian A.~C., Rees M.~J., 2008, MNRAS, 384, L19. doi:10.1111/j.1745-3933.2007.00413.x
\bibitem[\protect\citeauthoryear{Blandford \& K{\"o}nigl}{1979}]{1979ApJ...232...34B} Blandford R.~D., K{\"o}nigl A., 1979, ApJ, 232, 34. doi:10.1086/157262
\bibitem[\protect\citeauthoryear{Bloom \& Marscher}{1996}]{1996ApJ...461..657B} Bloom S.~D., Marscher A.~P., 1996, ApJ, 461, 657. doi:10.1086/177092
\bibitem[\protect\citeauthoryear{B{\"o}ttcher \& Dermer}{2010}]{2010ApJ...711..445B} B{\"o}ttcher M., Dermer C.~D., 2010, ApJ, 711, 445. doi:10.1088/0004-637X/711/1/445
\bibitem[\protect\citeauthoryear{B{\"o}ttcher et al.}{2013}]{2013ApJ...768...54B} B{\"o}ttcher M., Reimer A., Sweeney K., Prakash A., 2013, ApJ, 768, 54. doi:10.1088/0004-637X/768/1/54
\bibitem[\protect\citeauthoryear{Boutelier, Henri, \& Petrucci}{2008}]{2008MNRAS.390L..73B} Boutelier T., Henri G., Petrucci P.-O., 2008, MNRAS, 390, L73. doi:10.1111/j.1745-3933.2008.00539.x
\bibitem[\protect\citeauthoryear{Celotti \& Ghisellini}{2008}]{2008MNRAS.385..283C} Celotti A., Ghisellini G., 2008, MNRAS, 385, 283. doi:10.1111/j.1365-2966.2007.12758.x
\bibitem[\protect\citeauthoryear{Chiaberge \& Ghisellini}{1999}]{1999MNRAS.306..551C} Chiaberge M., Ghisellini G., 1999, MNRAS, 306, 551. doi:10.1046/j.1365-8711.1999.02538.x
\bibitem[\protect\citeauthoryear{Chadwick et al.}{1999}]{1999ApJ...513..161C} Chadwick P.~M., Lyons K., McComb T.~J.~L., Orford K.~J., Osborne J.~L., Rayner S.~M., Shaw S.~E., et al., 1999, ApJ, 513, 161. doi:10.1086/306862
\bibitem[\protect\citeauthoryear{Chatterjee et al.}{2019}]{2019MNRAS.490.2200C} Chatterjee K., Liska M., Tchekhovskoy A., Markoff S.~B., 2019, MNRAS, 490, 2200. doi:10.1093/mnras/stz2626
\bibitem[\protect\citeauthoryear{Chen et al.}{2013}]{2013A&A...553A.107C} Chen X., Rachen J.~P., L{\'o}pez-Caniego M., Dickinson C., Pearson T.~J., Fuhrmann L., Krichbaum T.~P., et al., 2013, A\&A, 553, A107. doi:10.1051/0004-6361/201220517
\bibitem[\protect\citeauthoryear{Cleary et al.}{2007}]{2007ApJ...660..117C} Cleary K., Lawrence C.~R., Marshall J.~A., Hao L., Meier D., 2007, ApJ, 660, 117. doi:10.1086/511969
\bibitem[\protect\citeauthoryear{Dermer \& Schlickeiser}{1993}]{1993ApJ...416..458D} Dermer C.~D., Schlickeiser R., 1993, ApJ, 416, 458. doi:10.1086/173251
\bibitem[\protect\citeauthoryear{Dermer \& Menon}{2009}]{2009herb.book.....D} Dermer C.~D., Menon G., 2009, herb.book
\bibitem[\protect\citeauthoryear{Finke, Dermer, \& B{\"o}ttcher}{2008}]{2008ApJ...686..181F} Finke J.~D., Dermer C.~D., B{\"o}ttcher M., 2008, ApJ, 686, 181. doi:10.1086/590900
\bibitem[\protect\citeauthoryear{Foschini et al.}{2007}]{2007ApJ...657L..81F} Foschini L., Ghisellini G., Tavecchio F., Treves A., Maraschi L., Gliozzi M., Raiteri C.~M., et al., 2007, ApJL, 657, L81. doi:10.1086/513271
\bibitem[\protect\citeauthoryear{Fossati et al.}{1998}]{1998MNRAS.299..433F} Fossati G., Maraschi L., Celotti A., Comastri A., Ghisellini G., 1998, MNRAS, 299, 433. doi:10.1046/j.1365-8711.1998.01828.x
\bibitem[\protect\citeauthoryear{Gao, Pohl, \& Winter}{2017}]{2017ApJ...843..109G} Gao S., Pohl M., Winter W., 2017, ApJ, 843, 109. doi:10.3847/1538-4357/aa7754
\bibitem[\protect\citeauthoryear{Ghisellini \& Tavecchio}{2008}]{2008MNRAS.386L..28G} Ghisellini G., Tavecchio F., 2008, MNRAS, 386, L28. doi:10.1111/j.1745-3933.2008.00454.x
\bibitem[\protect\citeauthoryear{Ghisellini \& Tavecchio}{2008}]{2008MNRAS.387.1669G} Ghisellini G., Tavecchio F., 2008, MNRAS, 387, 1669. doi:10.1111/j.1365-2966.2008.13360.x
\bibitem[\protect\citeauthoryear{Ghisellini \& Tavecchio}{2009}]{2009MNRAS.397..985G} Ghisellini G., Tavecchio F., 2009, MNRAS, 397, 985. doi:10.1111/j.1365-2966.2009.15007.x
\bibitem[\protect\citeauthoryear{Giannios}{2013}]{2013MNRAS.431..355G} Giannios D., 2013, MNRAS, 431, 355. doi:10.1093/mnras/stt167
\bibitem[\protect\citeauthoryear{Giannios, Uzdensky, \& Begelman}{2009}]{2009MNRAS.395L..29G} Giannios D., Uzdensky D.~A., Begelman M.~C., 2009, MNRAS, 395, L29. doi:10.1111/j.1745-3933.2009.00635.x
\bibitem[\protect\citeauthoryear{Griffiths et al.}{1979}]{1979ApJ...234..810G} Griffiths R.~E., Tapia S., Briel U., Chaisson L., 1979, ApJ, 234, 810. doi:10.1086/157560
\bibitem[\protect\citeauthoryear{Hayashida et al.}{2012}]{2012ApJ...754..114H} Hayashida M., Madejski G.~M., Nalewajko K., Sikora M., Wehrle A.~E., Ogle P., Collmar W., et al., 2012, ApJ, 754, 114. doi:10.1088/0004-637X/754/2/11
\bibitem[\protect\citeauthoryear{Hayashida et al.}{2015}]{2015ApJ...807...79H} Hayashida M., Nalewajko K., Madejski G.~M., Sikora M., Itoh R., Ajello M., Blandford R.~D., et al., 2015, ApJ, 807, 79. doi:10.1088/0004-637X/807/1/79
\bibitem[\protect\citeauthoryear{H.~E.~S.~S. Collaboration et al.}{2012}]{2012A&A...539A.149H} H.~E.~S.~S. Collaboration, Abramowski A., Acero F., Aharonian F., Akhperjanian A.~G., Anton G., Balzer A., et al., 2012, A\&A, 539, A149. doi:10.1051/0004-6361/201117509
\bibitem[\protect\citeauthoryear{Impiombato et al.}{2011}]{2011ApJS..192...12I} Impiombato D., Covino S., Treves A., Foschini L., Pian E., Tosti G., Fugazza D., et al., 2011, ApJS, 192, 12. doi:10.1088/0067-0049/192/1/12
\bibitem[\protect\citeauthoryear{Inoue \& Takahara}{1996}]{1996ApJ...463..555I} Inoue S., Takahara F., 1996, ApJ, 463, 555. doi:10.1086/177270
\bibitem[\protect\citeauthoryear{Jones, O'Dell, \& Stein}{1974}]{1974ApJ...188..353J} Jones T.~W., O'Dell S.~L., Stein W.~A., 1974, ApJ, 188, 353. doi:10.1086/152724
\bibitem[\protect\citeauthoryear{Katarzy{\'n}ski et al.}{2008}]{2008MNRAS.390..371K} Katarzy{\'n}ski K., Lenain J.-P., Zech A., Boisson C., Sol H., 2008, MNRAS, 390, 371. doi:10.1111/j.1365-2966.2008.13753.x
\bibitem[\protect\citeauthoryear{Katarzy{\'n}ski, Sol, \& Kus}{2001}]{2001A&A...367..809K} Katarzy{\'n}ski K., Sol H., Kus A., 2001, A\&A, 367, 809. doi:10.1051/0004-6361:20000538
\bibitem[\protect\citeauthoryear{Katarzy{\'n}ski, Sol, \& Kus}{2003}]{2003A&A...410..101K} Katarzy{\'n}ski K., Sol H., Kus A., 2003, A\&A, 410, 101. doi:10.1051/0004-6361:20031245
\bibitem[\protect\citeauthoryear{Komissarov et al.}{2007}]{2007MNRAS.380...51K} Komissarov S.~S., Barkov M.~V., Vlahakis N., K{\"o}nigl A., 2007, MNRAS, 380, 51. doi:10.1111/j.1365-2966.2007.12050.x
\bibitem[\protect\citeauthoryear{Li et al.}{2022}]{2022A&A...659A.184L} Li W.-J., Xue R., Long G.-B., Wang Z.-R., Nagataki S., Yan D.-H., Wang J.-C., 2022, A\&A, 659, A184. doi:10.1051/0004-6361/202142051
\bibitem[\protect\citeauthoryear{Liu et al.}{2023}]{2023MNRAS.526.5054L} Liu R.-Y., Xue R., Wang Z.-R., Tan H.-B., B{\"o}ttcher M., 2023, MNRAS, 526, 5054. doi:10.1093/mnras/stad2911
\bibitem[\protect\citeauthoryear{Lyutikov \& Lister}{2010}]{2010ApJ...722..197L} Lyutikov M., Lister M., 2010, ApJ, 722, 197. doi:10.1088/0004-637X/722/1/197
\bibitem[\protect\citeauthoryear{Marscher \& Gear}{1985}]{1985ApJ...298..114M} Marscher A.~P., Gear W.~K., 1985, ApJ, 298, 114. doi:10.1086/163592
\bibitem[\protect\citeauthoryear{Mastichiadis \& Kirk}{1997}]{1997A&A...320...19M} Mastichiadis A., Kirk J.~G., 1997, A\&A, 320, 19. doi:10.48550/arXiv.astro-ph/9610058
\bibitem[\protect\citeauthoryear{Nalewajko et al.}{2011}]{2011MNRAS.413..333N} Nalewajko K., Giannios D., Begelman M.~C., Uzdensky D.~A., Sikora M., 2011, MNRAS, 413, 333. doi:10.1111/j.1365-2966.2010.18140.x
\bibitem[\protect\citeauthoryear{Ostorero, Villata, \& Raiteri}{2004}]{2004A&A...419..913O} Ostorero L., Villata M., Raiteri C.~M., 2004, A\&A, 419, 913. doi:10.1051/0004-6361:20035813
\bibitem[\protect\citeauthoryear{Petropoulou, Giannios, \& Sironi}{2016}]{2016MNRAS.462.3325P} Petropoulou M., Giannios D., Sironi L., 2016, MNRAS, 462, 3325. doi:10.1093/mnras/stw1832
\bibitem[\protect\citeauthoryear{Petropoulou, Psarras, \& Giannios}{2023}]{2023MNRAS.518.2719P} Petropoulou M., Psarras F., Giannios D., 2023, MNRAS, 518, 2719. doi:10.1093/mnras/stac3190
\bibitem[\protect\citeauthoryear{Rani et al.}{2013}]{2013A&A...557A..71R} Rani B., Lott B., Krichbaum T.~P., Fuhrmann L., Zensus J.~A., 2013, A\&A, 557, A71. doi:10.1051/0004-6361/201321440
\bibitem[\protect\citeauthoryear{Schwartz et al.}{1979}]{1979ApJ...229L..53S} Schwartz D.~A., Doxsey R.~E., Griffiths R.~E., Johnston M.~D., Schwarz J., 1979, ApJL, 229, L53. doi:10.1086/182929
\bibitem[\protect\citeauthoryear{Sikora, Begelman, \& Rees}{1994}]{1994ApJ...421..153S} Sikora M., Begelman M.~C., Rees M.~J., 1994, ApJ, 421, 153. doi:10.1086/173633
\bibitem[\protect\citeauthoryear{Sironi, Petropoulou, \& Giannios}{2015}]{2015MNRAS.450..183S} Sironi L., Petropoulou M., Giannios D., 2015, MNRAS, 450, 183. doi:10.1093/mnras/stv641
\bibitem[\protect\citeauthoryear{Spada et al.}{2001}]{2001MNRAS.325.1559S} Spada M., Ghisellini G., Lazzati D., Celotti A., 2001, MNRAS, 325, 1559. doi:10.1046/j.1365-8711.2001.04557.x
\bibitem[\protect\citeauthoryear{Summerlin \& Baring}{2012}]{SB12} Summerlin, E. J., \& Baring, M. G., 2012, ApJ, 745, 63. doi:10.1088/0004-637X/745/1/63
\bibitem[\protect\citeauthoryear{Tavecchio \& Ghisellini}{2008}]{2008MNRAS.386..945T} Tavecchio F., Ghisellini G., 2008, MNRAS, 386, 945. doi:10.1111/j.1365-2966.2008.13072.x
\bibitem[\protect\citeauthoryear{Tavecchio, Maraschi, \& Ghisellini}{1998}]{1998ApJ...509..608T} Tavecchio F., Maraschi L., Ghisellini G., 1998, ApJ, 509, 608. doi:10.1086/306526
\bibitem[\protect\citeauthoryear{Urry \& Padovani}{1995}]{1995PASP..107..803U} Urry C.~M., Padovani P., 1995, PASP, 107, 803. doi:10.1086/133630
\bibitem[\protect\citeauthoryear{Vestrand, Stacy, \& Sreekumar}{1995}]{1995ApJ...454L..93V} Vestrand W.~T., Stacy J.~G., Sreekumar P., 1995, ApJL, 454, L93. doi:10.1086/309790
\bibitem[\protect\citeauthoryear{Wagner \& Witzel}{1995}]{1995ARA&A..33..163W} Wagner S.~J., Witzel A., 1995, ARA\&A, 33, 163. doi:10.1146/annurev.aa.33.090195.001115
\bibitem[\protect\citeauthoryear{Wang et al.}{2022}]{2022PhRvD.105b3005W} Wang Z.-R., Liu R.-Y., Petropoulou M., Oikonomou F., Xue R., Wang X.-Y., 2022, PhRvD, 105, 023005. doi:10.1103/PhysRevD.105.023005
\bibitem[\protect\citeauthoryear{Xue \& Cui}{2005}]{2005ApJ...622..160X} Xue Y., Cui W., 2005, ApJ, 622, 160. doi:10.1086/427933
\end{thebibliography}







\bsp	
\label{lastpage}
\end{document}